\documentclass{emulateapj}   

\usepackage{wrapfig}
\usepackage{graphicx}
\usepackage{epsfig}
\usepackage{epstopdf}

\newcommand{\Mach}{\mathcal{M}}

\def\yesfig{1}

\def\figver{1}

\usepackage{color}

\newcommand{\dofig}[1]{
\ifx\figver\yesfig
{#1}
\fi
}

\begin{document}

\title {Galaxy Cluster Radio Relics in Adaptive Mesh Refinement Cosmological Simulations: Relic Properties and Scaling Relationships}

\author{Samuel~W.~Skillman\altaffilmark{1,2}, Eric~J.~Hallman\altaffilmark{1,3,4},  Brian~W.~O'Shea\altaffilmark{5,6}, \\ Jack~O.~Burns\altaffilmark{1,7}, Britton~D.~Smith\altaffilmark{1,5}, Matthew J. Turk\altaffilmark{8,9}}

\altaffiltext{1}{Center for Astrophysics and Space Astronomy, Department of Astrophysical \& Planetary Science, University of Colorado, Boulder, CO 80309}
\altaffiltext{2}{DOE Computational Science Graduate Fellow}
\altaffiltext{3}{National Science Foundation Astronomy and Astrophysics Postdoctoral Fellow}
\altaffiltext{4}{Institute for Theory and Computation, Harvard-Smithsonian Center for Astrophysics, Cambridge, MA 02138}
\altaffiltext{5}{Department of Physics \& Astronomy, Michigan State University, East Lansing, MI, 48824}
\altaffiltext{6}{Lyman Briggs College and Institute for Cyber-Enabled Research, Michigan State University, East Lansing, MI, 48824}
\altaffiltext{7}{NASA Lunar Science Institute, NASA Ames Research Center, Moffet Field, CA, 94035}
\altaffiltext{8}{Columbia University, Department of Astronomy, New York,NY, 10025, USA}
\altaffiltext{9}{NSF Office of Cyberinfrastructure Postdoctoral Fellowship}

\email{samuel.skillman@colorado.edu}

\begin{abstract}
  Cosmological shocks are a critical part of large-scale structure
  formation, and are responsible for heating the intracluster medium
  in galaxy clusters.  In addition, they are also capable of
  accelerating non-thermal electrons and protons.  In this work, we
  focus on the acceleration of electrons at shock fronts, which is
  thought to be responsible for radio relics - extended radio features
  in the vicinity of merging galaxy clusters.  By combining high
  resolution AMR/N-body cosmological simulations with an accurate
  shock finding algorithm and a model for electron acceleration, we
  calculate the expected synchrotron emission resulting from
  cosmological structure formation.  We produce synthetic radio maps
  of a large sample of galaxy clusters and present luminosity
  functions and scaling relationships. With upcoming long wavelength
  radio telescopes, we expect to see an abundance of radio emission
  associated with merger shocks in the intracluster medium.  By
  producing observationally motivated statistics, we provide
  predictions that can be compared with observations to further improve our
  understanding of magnetic fields and electron shock acceleration.

\end{abstract}


\section{Introduction}

The assembly history of galaxy clusters are wrought with violent
mergers, high Mach-number flows, and extreme plasma physical
interactions.  Much of this results from a wide range of cosmological
structure formation shocks~\citep{Miniati:2001aa, Ryu:2003aa,
  Pfrommer:2006aa, Kang:2007ac, Hoeft:2008aa, Skillman:2008aa,
  Vazza:2009aa, Paul:2011aa}.  These shocks, however, do more than
simply heat the inflowing plasma.  They also accelerate electrons and
ions to relativistic speeds \citep{Bell:1978aa, Blandford:1978aa}.
See \citet{Drury:1983aa}, \citet{Blandford:1987aa}, and
\citet{Jones:1991aa} for reviews.  These relativistic particles then
act as signatures of merger and shock activity.  The relativistic
protons have radiative loss (e.g. collisions, pion decay, inverse
Compton scattering) times comparable to the Hubble time, and therefore
remain in the intracluster medium and contribute to the total pressure
of the gas \citep{Miniati:2001aa, Pfrommer:2006aa}.  On the other
hand, relativistic electrons have relatively short lifetimes, on the
order of a few hundred million years, and spend the remaining part of
their life emitting synchrotron radiation as they gyrate about
magnetic field lines.

Relativistic protons in the intracluster medium are, in principle, most
easily observed through their collisional interactions with thermal
ions, leading to pion decays that end in gamma-ray emission
\citep{Pfrommer:2006aa}.  This emission is being studied with the Fermi
satellite, and while preliminary results hint at low levels of
relativistic ions, long integrations of individual clusters are still
forthcoming \citep{Aleksic:2010aa}.

Relativistic electrons have been more extensively studied in several
galaxy clusters through their synchrotron radiation
~\citep{Rottgering:1997aa, Orru:2007aa, Bonafede:2009aa,
  van-Weeren:2009aa}.  These electrons are most frequently associated
with objects called radio $relics$ \citep{Ensslin:1998aa}, which have extended radio emission
in the cluster exterior, are associated with shocks, and have 
moderately polarized radio emission with spectral indices of $\alpha\approx1-2$ for 
surface brightness $S \sim \nu^{-\alpha}$).  They are also, by
definition, not associated with active galactic nuclei (AGN).  This
spectral shape most likely indicates that these electrons were
recently shock-accelerated \citep{Blandford:1987aa}.  Radio $halos$,
on the other hand, have low polarization and usually follow the X-ray
morphology in the centers of clusters.  They are thought to be
associated with turbulent acceleration and/or older populations of
previously shock-accelerated electrons
\citep[e.g.][]{Brunetti:2001aa}.

The origin of the shock-accelerated electrons is believed to be
primarily due to diffusive shock acceleration (DSA), as described in
~\citet{Blandford:1987aa}.  This is a first-order Fermi mechanism in which
electrons are accelerated by reflecting off magnetic field perturbations
created by plasma effects in shock waves.  Recent numerical studies by
\citet{Spitkovsky:2008aa} have shown success in reproducing this
mechanism through ab-initio simulations using particle-in-cell (PIC)
methods, though the incoming plasma flow was at a much higher velocity
than the shocks discussed here.  Studies of non-relativistic flow
are ongoing because of the difficulty associated with the range in
relevant timescales in such simulations \citep{Spitkovsky:2008aa}.

If these electrons are capable of reaching high enough energies, they
will emit synchrotron radiation in the presence of magnetic fields.  It
is widely believed that the cluster surroundings are magnetized at
relatively low field strengths, on the order of microgauss in the
ICM \citep{Fusco-Femiano:2001aa, Rephaeli:2003aa, Rephaeli:2006aa,
  Ryu:2008aa}, although measurements made by Faraday rotation indicate
$\sim10\times$ larger values \citep{Govoni:2006aa, Clarke:2001aa}.  Because the
shock-accelerated electrons are expected to have a power-law
distribution in energy, the synchrotron emission will also have a
power-distribution in frequency. There have been a number of recent
studies of these radio \textit{relics} in interferometric observations of
nearby galaxy clusters~\citep{Rottgering:1997aa,Orru:2007aa,
  Bonafede:2009aa,van-Weeren:2009aa}.  However, the total number of
clusters with known relics is still fairly small
($\sim22$) \citep{van-Weeren:2009aa}.  This is in part due to the low
surface brightness of these extended sources.

Studies of these objects comes at a critical point in time with a number
of upcoming improvements in radio astronomy capabilities.  The VLA is
currently being upgraded to the ``Expanded VLA'' (EVLA), which will be
roughly a factor of 10 better in terms of surface brightness sensitivity
due primarily to an increase in bandwidth of up to $1~\mathrm{GHz}$ at
1.5~GHz \citep{Napier:2006aa}.  Several other telescopes will be coming
online in the near future, such as LOFAR, possibly the SKA, and
even lunar farside low frequency arrays \citep{Burns:2009aa} allowing an
unprecedented view of the synchrotron Universe \citep{Rudnick:2009aa}.
Given the low surface brightness and spectrum of the emission, low
frequency observations ($\sim 1~\mathrm{GHz}$ and below) are most effective in
providing information on the radio relics and halos.

Here we set out to model these radio relics using high resolution,
adaptive mesh refinement (AMR) cosmological simulations.  These
simulations include both dark matter and adiabatic baryonic physics, and
allow us to model the shock acceleration of electrons and produce
observationally-relevant radio luminosity functions and scaling
relationships between cluster parameters such as synchrotron power,
mass, and X-ray luminosity.  By using large simulation volumes, we are
able to show statistics for thousands of objects, from which individual
morphological and evolutionary analyses can be carried out.

In Section \ref{sec:methods}, we introduce our simulation and galaxy
cluster sample set and describe our shock finding method and
synchrotron emission models. In Section \ref{sec:global_properties}, we
study the global statistics of radio relics through projections and
phase diagrams.  Then, in Section \ref{sec:individual_object_properties}, we
describe the properties of individual halos, and use their statistics
to generate radio luminosity scaling relationships and luminosity
functions.  We end with Section \ref{sec:discussion} where we discuss
the implications for future surveys, the limitations of our model, and
future directions.

\section{Methods}\label{sec:methods}

\dofig{
\begin{figure}[htp]
\centering
\includegraphics[width=0.45\textwidth]{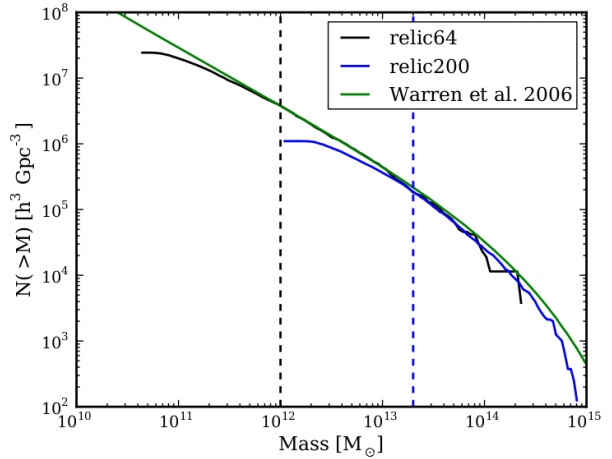}

\caption{Mass function of halos in \textit{relic64} and
  \textit{relic200} at $z=0$.  Halos are found using HOP with a
  minimum number of 30 dark matter particles.  Conservative estimates
  of the low mass cutoff are $10^{12}M_{\odot}$ and
  $2\times\sim10^{13}M_{\odot}$ for \textit{relic64} and
  \textit{relic200}, respectively.  This corresponds to 512 and 320
  particles.  The discrepancy at low mass for \textit{relic200} is
  because of the lack of resolution of low-mass halos and poor force
  resolution at small scales. Dashed lines show the lower limit for
  the resolved halos for \textit{relic64} and \textit{relic200} in
  black and blue, respectively.  Also shown are the fits from
  \citet{Warren:2006aa}.  Because of the similarity between the two
  cosmologies, these fits differ by less than thickness of the line in
  this mass range.}\label{fig:mass_func}
\end{figure}
}

\subsection{\textit{Enzo}}
All simulations were run using the \textit{Enzo} cosmology code
\citep{Bryan:1997aa,Bryan:1997ab, Norman:1999aa,
  OShea:2005aa, OShea:2005ab}.  While a full description can be found in the
cited papers, we will review the key aspects that are of importance to
this work.

\textit{Enzo} uses block-structured adaptive mesh refinement
 \citep[AMR;][]{Berger:1989aa} as a base upon which it couples an
Eulerian hydrodynamic solver for the gas with an N-Body particle mesh
(PM) solver \citep{Efstathiou:1985aa, Hockney:1988aa} for the
dark matter.  Users have the choice of solving the hydrodynamics with
several methods including the piecewise parabolic method
\citep[PPM;][]{Woodward:1984aa} extended for cosmological applications
by \citet{Bryan:1995aa} and the ZEUS finite-difference method
\citep{Stone:1992aa,Stone:1992ab}.  In this work we utilize both
methods, and restrict our studies to adiabatic gas physics.

The AMR method that \textit{Enzo} uses breaks the simulation domain into
rectangular solid volumes called $grids$.  These $grids$ contain many
computational elements called $cells$ that set the resolution scale.
Each grid exists on a $level$ of refinement determined by the spatial resolution of its cells that ranges from $0-l_{max}$, where $l_{max}$ is
defined by the user.  Once $cells$ within a given $grid$ satisfy the
refinement criteria (based on overdensity, minimum resolution of the
Jeans length, local gradients of hydrodynamical quantities, shocks, or
cooling time), a new $grid$ is created at the next higher level.  In our
simulations, we refine on overdensity of the gas and dark matter fields.

\subsection{Simulations}\label{sec:simulations}
We will focus on two simulations that both use N-body dynamics for the
dark matter and adiabatic baryonic physics. The first simulation,
hereby denoted as \textit{relic64}, has a comoving volume of $(64\
h^{-1}\mathrm{Mpc})^3$ with $256^3$ root-grid cells and up to 6 levels
of additional refinement.  The AMR is done by inserting a
higher-resolution region wherever a cell satisfies the refinement
criteria.  Here we require a gas or dark matter overdensity ($\delta
\equiv \rho/\bar\rho$ where $\bar\rho$ is the average density of gas
or dark matter, respectively) of $8$ to refine.  Because refinement
effectively splits a cell into 8 cells, this ensures that cells on
each level have similar amounts of mass. This allows for a peak
spatial resolution of $3.9~h^{-1}\mathrm{kpc~(comoving)}$. The
simulation uses the ZEUS hydrodynamic solver, with initial conditions
from an \citet{Eisenstein:1999aa} power spectrum with a spectral index
$n_s=0.97$. The cosmological parameters used are $\Omega_M = 0.268$,
$\Omega_{B} = 0.0441$, $\Omega_{CDM} = 0.2239$ ,
$\Omega_{\Lambda}=0.732$, $h= H_0/(100\ \mathrm{km}\ \mathrm{s}^{-1}
\mathrm{Mpc}^{-1})=0.704$, and $\sigma_8 = 0.82$.  The dark matter
mass resolution is $1.96\times10^9~h^{-1}M_\odot$. The simulation was
started at a redshift of $z=99$ and run until $z=0$, using
approximately $300,000$ cpu-hours on the Texas Advanced Computing
Center (TACC) Ranger supercomputer.

We use a second simulation with a larger volume
$(200~h^{-1}~\mathrm{Mpc})^3$ with more modest resolution,
\textit{relic200}, to capture a higher mass range for our simulated
galaxy clusters. As before, it uses $256^3$ root-grid cells and up to
5 levels of AMR. It has a peak resolution of $24.4~ h^{-1}
\mathrm{kpc~(comoving)}$ and a dark matter mass resolution of
$6.23\times10^{10}~h^{-1}M_\odot$. The simulation uses a slightly
different cosmology of $n_s=0.96$, $\Omega_M = 0.279$, $\Omega_{B} =
0.046$, $\Omega_{CDM} = 0.2239$ , $\Omega_{\Lambda}=0.721$, $h=
H_0/(100\ \mathrm{km}\ \mathrm{s}^{-1} \mathrm{Mpc}^{-1})=0.701$, and
$\sigma_8 = 0.817$, consistent with WMAP Year-5 results
\citep{Komatsu:2008aa}. This simulation uses the PPM hydrodynamic
solver, was also run on the TACC Ranger, and took approximately
$100,000$ cpu-hours to complete.

The mass functions of the two simulations are shown in Figure
\ref{fig:mass_func}. To calculate the mass function we begin by
finding all the halos using the halo-finding algorithm HOP
\citet{Eisenstein:1998aa}, implemented in
$yt$\footnote{http://yt.enzotools.org}, an analysis and visualization
system written in Python, designed for use with the adaptive mesh
refinement codes including \textit{Enzo} \citep{Turk:2011aa}. This
method finds halos by ``hopping'' from one dark matter particle to its
most dense neighbor until a particle is its own highest density
neighbor.  All particles that find the same densest particle are then
grouped into a single halo.  The \textit{relic64} simulation contains
$10^{11}M_\odot-10^{14}M_\odot$ halos, but in our analysis we only
consider objects with masses above $10^{12}M_\odot$, corresponding to
$\sim510$ dark matter particles.  This run was primarily designed to
have superb resolution capable of capturing the morphology and
structure of the relics and shocks.  The \textit{relic200} simulation
contains $5\times10^{12}M_\odot-8\times10^{14}M_\odot$ halos. For this
simulation, we consider only halos above $2\times10^{13}M_\odot$,
corresponding to $\sim320$ dark matter particles. This simulation is designed
to study the statistics of medium-sized clusters.  While neither of
these simulations capture the most massive clusters in the Universe
(e.g. Coma), they provide insight to radio relic origins, structure,
and evolution.  Studies of very large volume simulations are reserved
for future work.

\subsection{Shock Finding}\label{sec:shock_finding}

To identify the shocks that ultimately accelerate the electrons
that emit synchrotron radiation, we need an accurate shock
identification algorithm.  For this we use the temperature-jump method
given in \citet{Skillman:2008aa}.  Here we present an overview of the method
for completeness.  We use the Rankine-Hugoniot temperature jump
conditions to derive the Mach number: 
\begin{equation}
\frac{T_2}{T_1} = \frac{(5\Mach^2 - 1)(\Mach^2 + 3)}{16\Mach^2},
\end{equation}
where $T_2$ and $T_1$ are the post-shock~(downstream) and
pre-shock~(upstream) temperatures, respectively. $\Mach$ is the Mach
number using the upstream (pre-shock) gas.

 A cell is determined to have a shock if it meets the following requirements: 
\begin{eqnarray}
\nabla \cdot \vec{v} < 0,\ \ \nabla T \cdot \nabla S > 0,\ \ T_2 > T_1,\ \ \rho_2 > \rho_1,
\end{eqnarray}
where $\vec{v}$ is the velocity field, $T$ is the temperature, $\rho$
is the density, and $S=T/\rho^{\gamma-1}$ is the entropy. In our
analysis, as in \citet{ Skillman:2008aa}, we have set a
minimum preshock temperature of $T=10^4$~K since the low-density gas in
our cosmological simulations is assumed to be ionized~(a reasonable
assumption at $z < 6$). Therefore, any time the pre-shock temperature
is lower than $10^4$~K, the Mach number is calculated from the ratio of
the post-shock temperature to $10^4$~K.

Once a shock is found, we identify the cell with the most negative
flow divergence, choosing from a ray aligned with the temperature
gradient.  Therefore, even if several cells in a row qualify as a
shock, only the ``center'' of the shock is marked as a shock, and the
temperature jump is taken from the full jump across all the shocked
cells.  This relieves problems when shocks are spread out over several
cells, especially when not aligned with the coordinate axes.  In
addition, this method has been implemented to run ``on-the-fly'' in
\textit{Enzo} so that post-processing of the data is not
needed.  This method then saves the Mach number and quantities such as
the pre-shock density and temperature directly along with the other
hydrodynamical quantities during simulation output.  The unique
feature of this shock-finder is its ability to accurately identify
off-axis shocks within AMR simulations and quantify their Mach number
even if the shock is identified as being spread out across several cells.

\dofig{
\begin{figure*}[htp]
  \centering
  \includegraphics[width=1.0\textwidth]{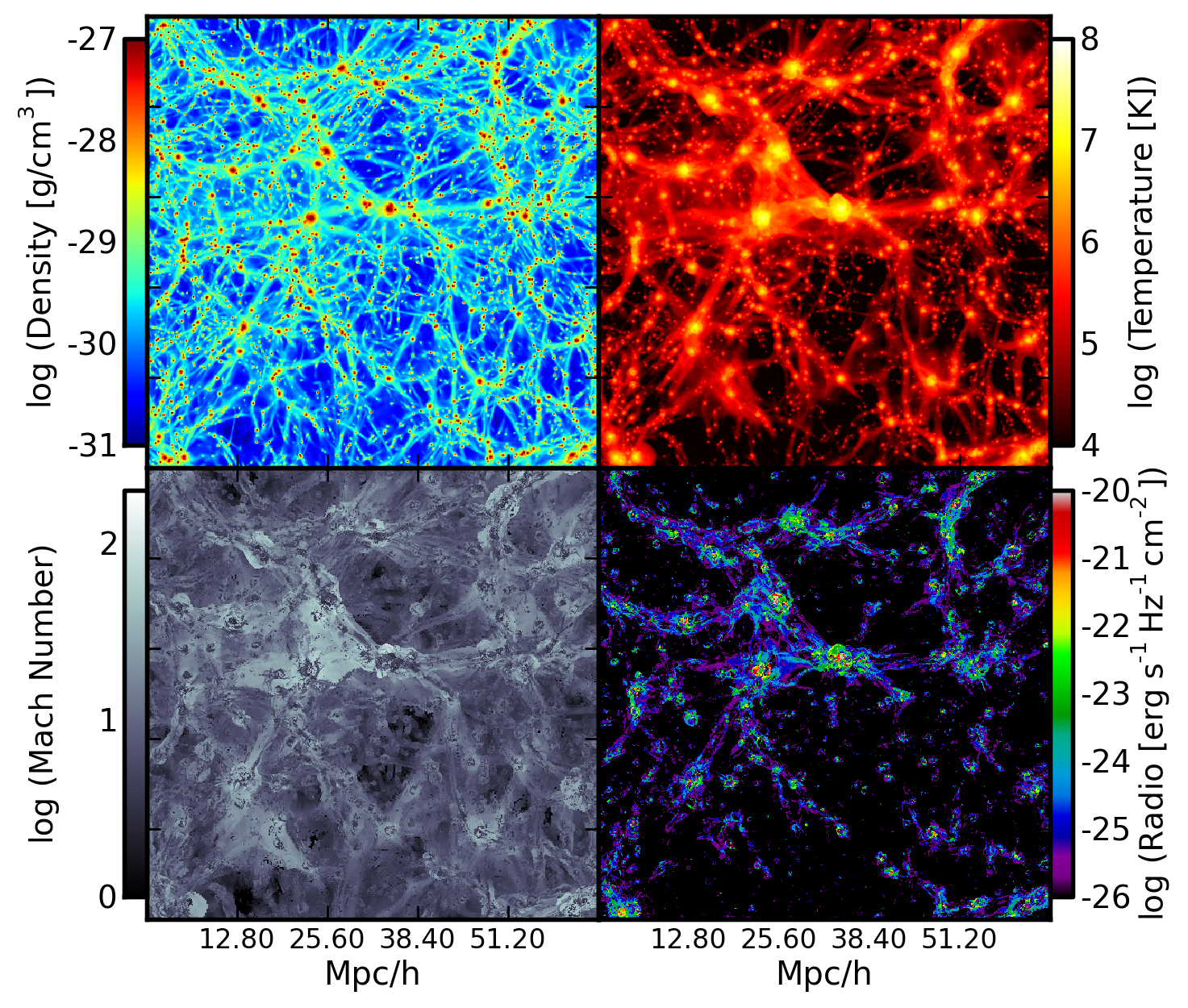}

  \caption{Projection of several quantities through the entire \textit{relic64} simulation volume at $z=0$. Shown are
    mass-weighted density [$\mathrm{g}\ \mathrm{cm}^{-3}$] (upper left), mass-weighted temperature [$\mathrm{K}$] (upper
    right), radio emission-weighted Mach number (lower left), and
    radio flux density [$\mathrm{erg}\ \mathrm{s}^{-1} \mathrm{Hz}^{-1}\mathrm{cm}^{-2}$] (lower
    right).}
  \label{full-box-64}
\end{figure*}
}

\dofig{
\begin{figure*}[htp]
  \centering
  \includegraphics[width=1.0\textwidth]{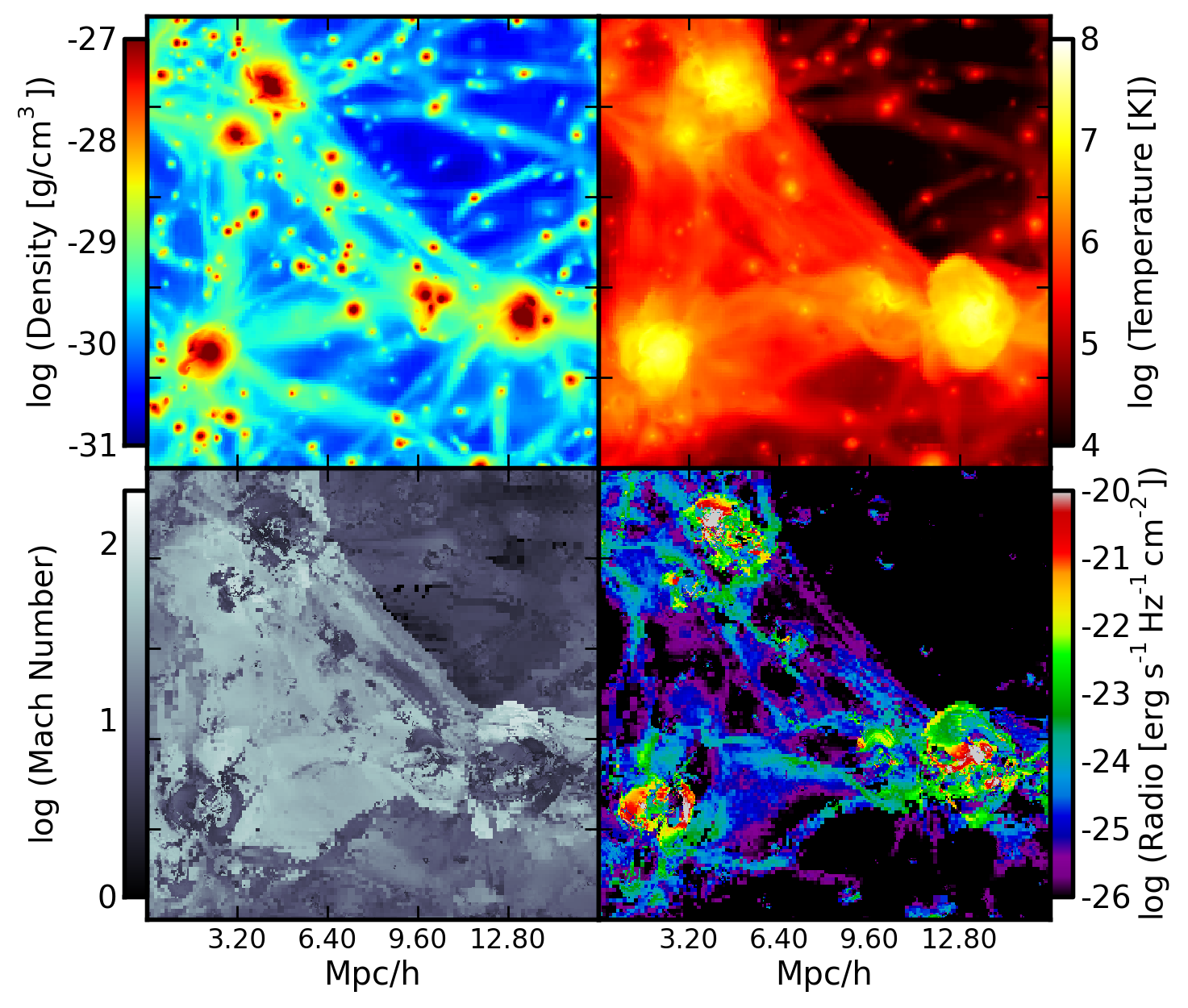}

  \caption{A zoom-in of Figure \ref{full-box-64}, now $16 \mathrm{Mpc/h}$ wide.}
  \label{zoom-box-64}
\end{figure*}
}

\subsection{Synchrotron Emission}\label{sec:synchrotron_emission}
In order to estimate the synchrotron emission from the shock waves, we
follow the method of \citet{Hoeft:2007aa}. Here we summarize the main
features of the model. The first assumption is that the electrons are
accelerated to a power-law distribution that is related to the Mach
number from diffusive shock acceleration theory.  These accelerated
electrons form an extension to the thermal, Maxwellian distribution that
has a power-law form and exponential cutoff related to balancing the
acceleration and cooling times of the electrons.

The accelerated electrons then emit in the radio through synchrotron
radiation. Since we are not performing magnetohydrodynamic
simulations, we assume that the magnetic field is governed by flux
freezing such that the magnetic field strength is related to the
density by $B\ = 0.1 \mu G (\frac{n}{10^{-4}\mathrm{cm}^{-3}})^{2/3}$,
where $n$ is the number density.  This is a reasonable assumption even
in merging clusters, as was found by \citet{Roettiger:1999aa}.  In
Section \ref{sec:luminosity_function} and Appendix \ref{sec:append_emission} we explore
the effect of a modified magnetic field model.

The total radio power from a shock wave of area $A$, frequency
$\nu_{obs}$, magnetic field $B$, electron acceleration efficiency
$\xi_e$, electron power-law index $s$ ($n_e \propto E^{-s}$), post-shock electron density
$n_e$ and temperature $T_2$ is \citep{Hoeft:2007aa} 
\begin{eqnarray}
  \label{eq:2}
\frac{dP(\nu_{obs})}{d\nu} = 6.4\times10^{34}erg\ s^{-1}\ Hz^{-1} \frac{A}{\mathrm{Mpc}^2} \frac{n_e}{10^{-4} \mathrm{cm}^{-3}}\cr
\frac{\xi_e}{0.05}(\frac{\nu_{obs}}{1.4GHz})^{-s/2}  \times (\frac{T_2}{7 keV})^{3/2}\cr
\frac{(B/\mu G)^{1+(s/2)}}{ (B_{CMB}/\mu G)^2 + (B/\mu G)^2} \Psi (\Mach).  
\end{eqnarray}
Note that the radiation spectral index is related to the electron
spectral index by $\alpha=(s-1)/2$.  $B_{CMB}$ is defined as the
magnetic field corresponding to the energy density of the CMB.  It has
a value of $B\equiv3.47 \mathrm{\mu G}(1+z)^2$, and accounts for the
inverse Compton emission that is simultaneously cooling the electrons
along with their synchrotron emission.  The final term, $\Psi
(\Mach)$, is a dimensionless quantity that contains dependencies on
the shock Mach number such that at $\Psi(2.5) \sim 10^{-3}$ and
approaches $1$ for $\Mach > 10$. It can be thought of as a shape
factor that, together with $\xi$, defines the acceleration efficiency
as a function of Mach number.  Note that for calculations with $z>0$,
we modify $\nu_{obs}\to\nu_{obs}\times(1+z)$ since we observe the
redshifted emission.

In all of our analysis, we use $\xi_e=0.005$, which was found by
\citet{Hoeft:2008aa} to match radio emission in known relics to
similar mass clusters in their simulation.  While the true value of
this parameter is quite uncertain from current observational and
theoretical constraints, the relationship between it and the total
radio power is linear.  Therefore, if we underestimate the electron
acceleration efficiency by a factor of 10, it will lead to a derived
radio power that is low by a factor of 10, and thus all relationships
between radio power and other quantities (e.g. mass, x-ray luminosity)
simply need to be rescaled.  While we have chosen the dimensionless
shape factor $\Psi (\Mach)$ from \citet{Hoeft:2007aa}, there are still
uncertainties in the efficiency of acceleration as a function of Mach
number.  However, exploration of the effects of these uncertainties
are beyond the scope of this work.  We also ignore the effects of
re-accelerated $\gamma\sim200$ electrons from radio galaxies.

\dofig{
\begin{figure*}[htp]
  \centering
 \includegraphics[width=0.9\textwidth]{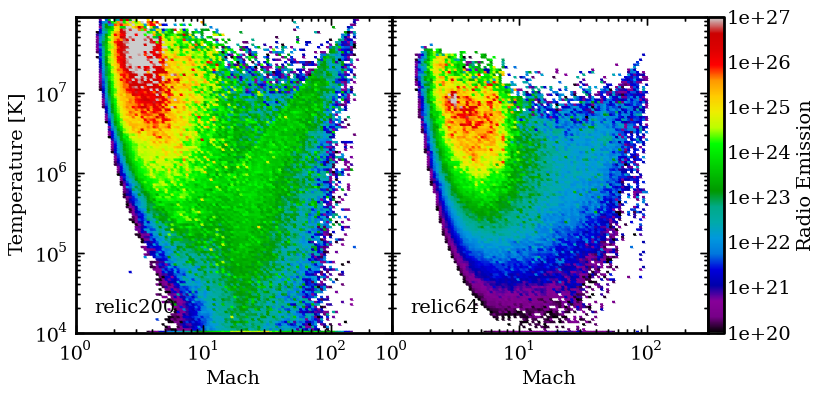}
\caption{2D Phase Diagram of the relic radio emission in
    temperature-Mach number space for the \textit{relic200}(left) and
    \textit{relic64}(right) simulation.  The temperature is that of the
    cell in the center of the shock. The color indicates the amount of
    1.4 GHz radio emission [$\mathrm{erg}~\mathrm{s}^{-1}
    \mathrm{Hz}^{-1}/(\mathrm{Mpc}/h)^3/(dlog\Mach~dlogT)$] for a given
    value of temperature and Mach number.}
  \label{fig:mach-temp-radio}
\end{figure*}
}

\section{Global Properties of Radio relics}\label{sec:global_properties}
\subsection{Full Box Projections}\label{sec:full_box_projections}

To begin our study of radio relics, we first performed simple
projections of the radio emission through the entire simulation
volume.  An example is shown in Figure \ref{full-box-64} along with
projections of density, temperature and Mach number.  For quantities
such as density, temperature and Mach number in an AMR simulation, we
choose to weight each cell by a secondary quantity since a simple
average along the line of sight for each cell would bias the most
highly refined regions because of their increased number of cells.
Therefore, we choose to weight the density and temperature fields by
cell mass, and the Mach number by the radio emission.  This has the
effect of pulling out the values of density and temperature from the
densest regions, and the shocks that contribute the most to the radio
emission.  For radio and X-ray fields, we project the emissivities
[energy/time/volume] without a weight, leading to final values with
units of [energy/time/area]. Mathematically, a weighted projection
(here along the z-axis) is defined by:
\begin{eqnarray}
  \label{eq:projection}
  P_z(x,y) = \frac{\int w(x,y,z) v(x,y,z)  dz}{\int w(x,y,z) dz}
\end{eqnarray}
where $w(x,y,z)$ is the weight quantity at that location and
$v(x,y,z)$ is the value of the projected quantity.  To evaluate this
integral in our AMR setting, the integral traverses the box along
cells that are at the highest refinement for a given point in space,
and ignores cells that are covered by more highly refined regions.
This, like the bulk of our analysis, is done using $yt$, detailed
above.

In Figures \ref{full-box-64}-\ref{zoom-box-64}, we see that in general
the radio emission traces out the large scale structure seen in the
density projection.  Additionally, the emission is also highly
correlated with the temperature structure.  However, the correlation
with Mach number is more interesting.  In the projection of Mach
number, we see shocks with strengths up to $\Mach \sim10-100$
throughout the volume in filaments and cluster edges, whereas the peak
radio emission only shows up in small, curved arcs within clusters.
At the location of these arcs, the value of the Mach number projection
drops to values between $\Mach \sim3-10$.  This shows that the
strongest shocks which are most likely external, accretion, shocks are not
responsible for the bright radio emission, and that it is instead the
interior shocks \citep{Ryu:2003aa, Skillman:2008aa}, as was found by
\citet{Hoeft:2008aa} with moderate strengths, that shine in the radio.
This can be understood by the fact that it is the mass flux of gas
through shocks that is most important since that determines the number
of electrons that can be accelerated.  Therefore, while the Mach
number is much lower for the interior shocks, the shock velocity stays
roughly constant while the pre-shock density is much higher, yielding
more accelerated electrons.  In the projection of Mach number, this
results in the appearance of ``veins'' lining the interior of the
filaments, ``arcs'' in the periphery of the clusters, and ``holes'' in
the centers of the clusters.  While they are decrements in the
projection of Mach number, they are the bright areas in the radio
emission.

The lack of strong emission in the accretion shocks suggests that
having a hot, dense plasma is more important than the Mach number of
the shock. This can be understood by Equation \ref{eq:2}.  Since
$\frac{dP}{d\nu}\propto A\ n_e \xi \nu T^{3/2}
\frac{B^{1+s/2}}{B_{CMB}^2 + B^2}$ and in most cluster situations
$B_{CMB}^2 > B^2$ and $s = 2\alpha + 1 ~= 3$, we have that
$\frac{dP}{d\nu}\propto n_e B^{5/2} \propto n_e (n_e^{2/3})^{5/2}
\propto n_e^{8/3}$.  This implies that since the density in the
accretion shocks is $\approx 10^2 - 10^3$ times lower than that in a
merger shock, the power emitted will be down by a factor of $\approx
2\times 10^3 - 10^5$.  Therefore the features we see observationally
are more likely to be related to merger shocks than accretion shocks.

\dofig{
\begin{figure*}[htp]
  \centering
  \includegraphics[width=0.9\textwidth]{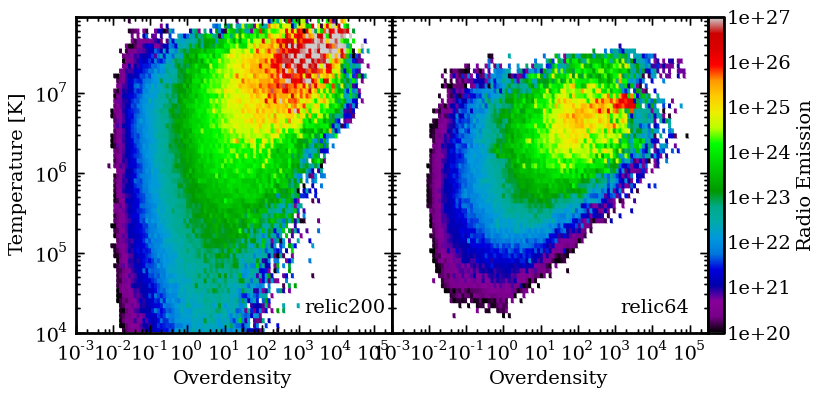}
 \caption{2D Phase Diagram of the relic radio emission in
    temperature-overdensity space for the \textit{relic200}(left) and
    \textit{relic64}(right) simulation. The color indicates the amount of 1.4
    GHz radio emission [$\mathrm{erg}~\mathrm{s}^{-1} \mathrm{Hz}^{-1}/(\mathrm{Mpc}/h)^3/(dlog\delta~dlogT)$] for a given
    value of temperature and overdensity.  }
  \label{fig:temp-od-radio}
\end{figure*}
}

\subsection{Phase Diagrams}\label{sec:phase}
The second method we use to study the bulk properties of the radio
emitting plasma is phase diagrams.  Such diagrams are the equivalent
of a two-dimensional histogram.  Here we use them to study the gas
properties of the radio emitting regions.

The structure of these diagrams is as follows.  For a given simulation
output, we construct x and y-axis bins that are equally spaced
logarithmically in two fields.  Within each of these 2D bins, we
integrate the total amount of a given quantity such as radio emission.
This integrated value is normalized by the comoving volume of the
simulation in order to give a comparable value between different
physical size simulations.  We have found three particularly insightful
quantities to examine in a range of permutions: temperature,
overdensity, and Mach number.

We have constructed one such phase diagram, seen in Figure
\ref{fig:mach-temp-radio}, in which the x-axis is the Mach number, the
y-axis is temperature, and the bins are colored by the total radio
emission in that bin.  The total integrated emission is normalized by
the volume of the simulation and the size of the bins.  As such, one
reads this figure as ``At Mach number $x$ and temperature $y$, there
is $z$ amount of radio emission per comoving Mpc/h per $\Delta
log\Mach \Delta log T$.''  The utility of these diagrams is
demonstrated in Figure \ref{fig:mach-temp-radio}, where it is
immediately clear that the bulk of the radio emission in both
simulations originates from hot gas with
$T=10^6-5\times10^7\mathrm{K}$, and Mach number $\Mach=3-10$.  This
reinforces our earlier hypothesis that the radio features are
generated from interior shocks associated with merging subclusters
that have low Mach numbers but high mass and energy flux due to the
high relative density, and therefore shock velocity, of cluster cores.
Second, it points out that shocks with $\Mach=20-100$ have little role
at $z=0$ in producing appreciable radio emission.  In fact, their
integrated luminosity is a factor of $500-1000$ less than their
low-Mach number counterparts.

At first glance, one also picks out a diagonal structure in the
\textit{relic200} phase diagram that seems to be an upper limit on the
temperature for a given Mach number.  This is a very interesting feature
that has a simple explanation.  We calculate our Mach number using a
minimum pre-shock temperature of $10^4\ \mathrm{K}$.  Now, while the gas
at the location of the shock is not necessarily the pre- or post-shock
temperature, it is bounded by those two values.  This is because the
shock location is based on the cell with the most convergent flow, not
the location of the pre- or post-shock gas.  Because of this, if gas
with pre-shock temperature $T_{1} < 10^4\ \mathrm{K}$ is being accreted,
the gas at the location of the shock will have a maximum temperature of
$T_{max} \le \frac{(5\Mach^2 - 1)(\Mach^2 + 3)}{16\Mach^2}10^4\mathrm{K}
$.  This maximum coincides perfectly with the diagonal feature.
Therefore, any gas below this line is likely pristine gas (that is, gas
that has not been previously shocked) being accreted onto filaments or
clusters for the first time.  This gives us a proxy for the relative
amount of accretion in a simulation.

We can then use this diagnostic to study the role of accretion in the
\textit{relic64} and \textit{relic200} simulations.  While
\textit{relic64} does have a small amount of accretion, it is far
below that of \textit{relic200}.  This is because of the different
mass clusters present in each of the two simulations;
\textit{relic200} has clusters that are up to an order of magnitude
more massive than in \textit{relic64}.  Recalling that the accretion radius
$r_s = \frac{G M}{c_s^2}$ scales with mass \citep{Bondi:1944aa}, the
clusters in \textit{relic200} are able to pull in and accrete more gas
than those in \textit{relic64}.  This behavior will be studied as a
function of redshift below.

As we did in the temperature-Mach number phase space, we now
examine the behavior in the temperature-overdensity plane in Figure
\ref{fig:temp-od-radio}, where overdensity is defined as
$\rho/\bar{\rho}$, where $\bar{\rho}$ is the mean matter density of
the Universe, $\Omega_M \rho_{crit}$.  Here our earlier findings are reinforced - the
strongest emission is coming from the densest, hottest regions in the
simulations.  

Both simulations exhibit the same general properties, though
\textit{relic200} has hotter gas again due to the larger clusters.
This suggests that not only will the most massive clusters likely be
associated with the strongest radio emission, but that the strongest
features will arise from merger shocks passing through the centers of
cluster, which seems to be the case observationally.  Below an
overdensity of $\sim10-30$, the radio emission greatly decreases.
This strongly disfavors the possibility of seeing cluster accretion
shocks in agreement with \citet{Hoeft:2008aa}.  This is compounded by
the fact that these accretion features are more diffuse and therefore
have reduced surface brightness compared to the more compact merger
shocks, making them difficult to study observationally.  If we then
compare \textit{relic64} to \textit{relic200}, we see that there is a
relative absence of accreting gas in the \textit{relic64} simulation,
reinforcing our earlier findings that \textit{relic64} has less
accretion due to the smaller mass halos compared to \textit{relic200}.
In the following section we study this accretion as an evolutionary
tracer in more depth.

\dofig{
\begin{figure*}[htpb]
  \centering
 \includegraphics[width=1.0\textwidth]{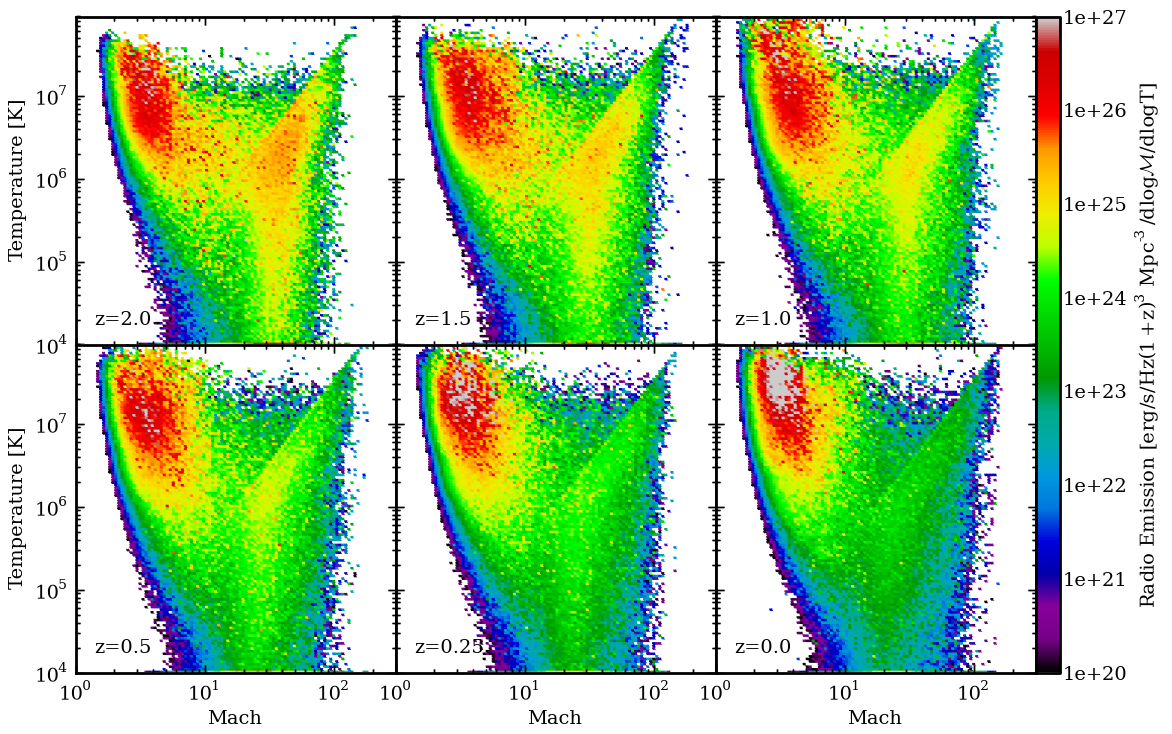}

  \caption{2d Phase Diagram of the relic radio emission in
    temperature-Mach number space for the \textit{relic200} simulation for
    varying redshifts.  From top left to bottom right, we have $z=2.0,
    1.5, 1.0, 0.5, 0.25, 0.00$}
  \label{fig:temp-mach-radio-red}
\end{figure*}
}

\subsection{Radio Emission as a Proxy for Cluster
  Accretion}\label{sec:radio_emission_accretion}

We have seen at $z=0$ that there is a relatively small amount of radio
emission coming from cluster and filament accretion shocks. We now
investigate whether or not this holds for earlier times. Figure
\ref{fig:temp-mach-radio-red} is an analog to our prior phase diagrams,
but we now show the evolution of this phase diagram for \textit{relic200} back
to $z=2.0$. We have normalized the emission by comoving volume in order
to avoid confusion with the expansion of space. Even with this taken
out, we see that there is a strong evolutionary trend in the origin of
the radio emission.

At $z=2$, we see that the emission from accretion shocks is comparable
to, if not above, that of the interior merger shocks. This roughly
translates to an equal amount of thermal energy being processed by
mergers and accretion shocks at $z=2$ since the efficiency of
acceleration does not differ dramatically between the two cases. The
line that we previously identified with accretion shocks is now quite
strong, and there is even a dominant population within the accretion
regime, centered around $10^6 \mathrm{K}$ and Mach numbers of $40-60$.

By $z=1$, the emission from accretion shocks has dropped by a factor
of 10 while the merger shocks have increased by a factor of
$2-3$. Finally, by $z=0$, nearly all of the radio emission due to
accretion shocks has disappeared while the interior shocks have
increased by a factor of 10 from $z=2$. This drop in radio emission
from accretion shocks coincides with the Universe beginning to
accelerate due to 'dark energy' at $z=0.75-1.0$. This process
manifests itself by both depleting the voids leading to less mass to
accrete, and decoupling of clusters from the cosmic
expansion. Therefore, instead of growing in mass, and growing the
radius of influence, dark energy dominates and pulls all the remaining
matter away from the cluster faster than it can grow.  We note here
that this assumes that there is no evolution in the relative magnetic
field strengths or acceleration efficiencies between merger and
accretion shocks.  It may be the case that the magnetic field strength
at accretion shock locations is lower or higher at early times
compared to the interior of proto-clusters.  However, since we do not
have conclusive evidence to this evolution, we have chosen to adopt
the simplest model that assumes no evolution.

This decrease in radio emission from accretion shocks is similar to the
depletion of cosmic ray proton acceleration at $z<1$ from accretion
shocks, found in \citet{Skillman:2008aa}.  This behavior is a novel
perspective to view the effects of Dark Energy.  If the radio
relics from these accretion shocks are observable in the future, one
should see a decrease in their frequency and power as $z$ decreases.  

\dofig{
\begin{figure*}[htp]
\centering
\includegraphics[width=0.99\textwidth]{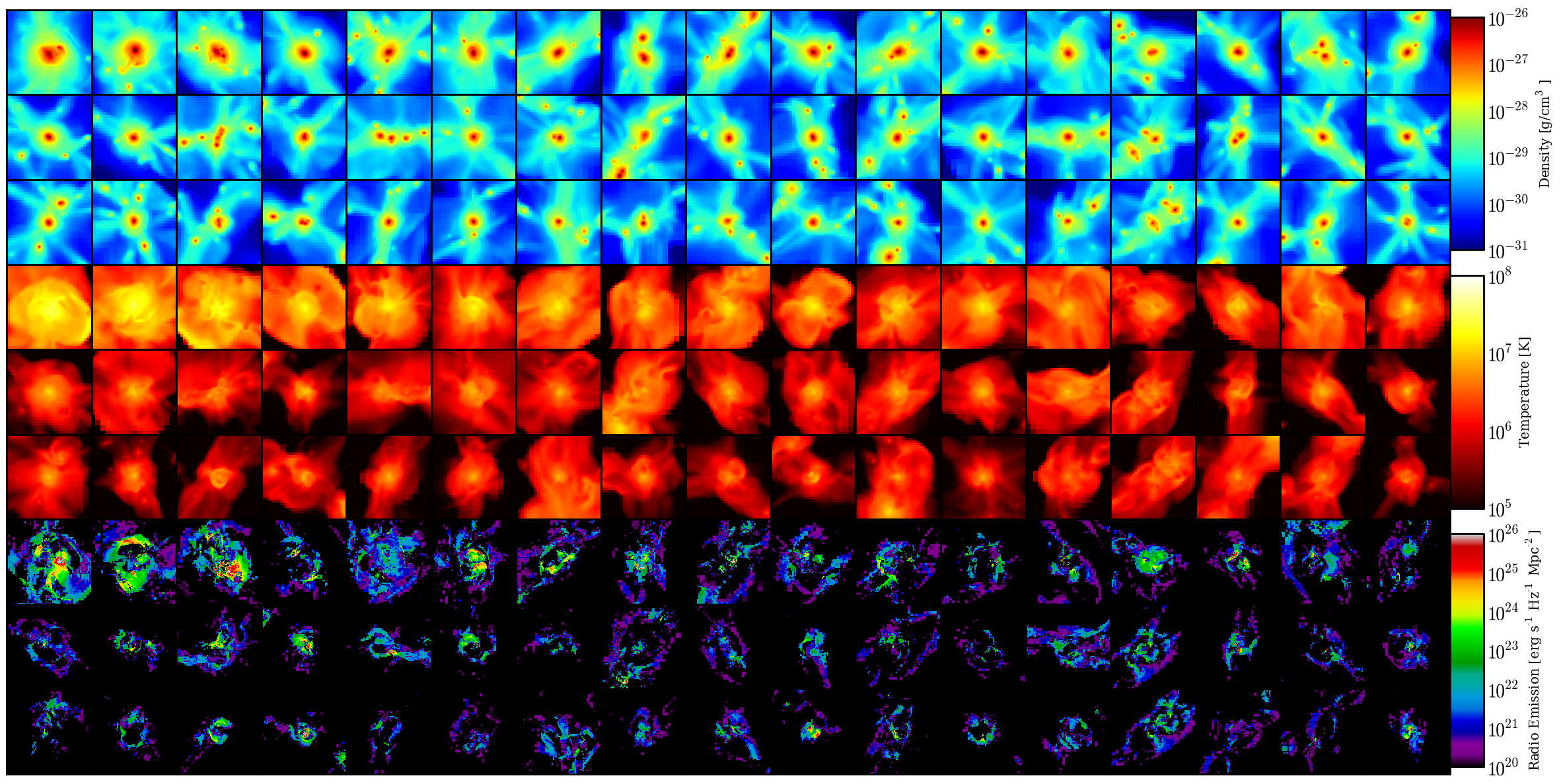}
\caption{ Density (top), temperature (middle) and 1.4GHz radio emission
  (bottom) for the 51 most massive halos in the \textit{relic64}
  simulation at $z=0$.  Mass decreases from top left to bottom
  right ($2.5\times10^{14} - 2.0\times10^{13} M_{\odot}$).  Each
  individual image is $4~\mathrm{Mpc}/h$ across.  All images are
  projections down the x-axis. }\label{fig:halos}
\end{figure*}
}

\section{Individual Object Properties}\label{sec:individual_object_properties}
\subsection{Cluster Projections}\label{sec:halo_projections}

In this section, we take the opposite approach from the previous section
and examine projections of individual objects. We begin with our list of
halos and make radial profiles that start at the density peak and
continue to the previously found $r_{200}$, defined here as the radius
where $\delta \equiv \rho/\bar{\rho} = 200$.  We call the mass enclosed
within this radius the $virial~mass$, and also record several other
quantities, such as X-ray luminosity and radio power within the virial
radius.

We begin by demonstrating the power of having a large sample of
clusters in a single simulation by projecting \textit{only} the 51
most massive clusters at $z=0$ along the x-axis in \textit{relic64} in
Figure \ref{fig:halos}.  The width and depth of each individual
projection here is 4 Mpc/h.

The most important result gleaned from these images is the
morphological properties of cluster structure.  If we first examine
the gas density (top), we see that while there is some amount of
substructure, the density is centrally concentrated.  Since X-ray
emission closely follows the density distribution of the gas, this
implies that the X-ray emission will be brightest in the centers of
clusters.  However, the radio emission (bottom) is brightest on the
edges of the clusters and has very little correlation with the density
structure.  Instead, it more closely follows the temperature structure
(middle).  This is because the temperature is more strongly affected
by shocks than the density (recall $\rho_2/\rho_1 \le 4$ from shock
jump conditions for $\gamma=5/3$).  Note, however, that the emission is still confined
within high density regions inside the virial radius.  Immediately from
these images, we expect radio emission to be anti-coincident with the
X-ray emission, as is seen in existing $relic$ examples
\citep{Giacintucci:2008aa, van-Weeren:2009ab, Bonafede:2009aa,
  Clarke:2006aa}.  This behavior implies that shocks are more likely
to appear in radio imaging than in X-ray surface brightness maps.

Also visible in the radio emission are common features such as arcs
and rings.  These features are due to merging subclusters as their bow
shocks propagate through the ICM.  These shapes are similar to what is
seen in observed radio relics.  This similarity supports our claim
that the morphology of these objects is related to the location of
shocks, as was originally suggested in \citet{Ensslin:1998aa}. In a
few rare situations (here in 2-3 clusters), these arcs appear in the
very center of the cluster.  Because the surrounding medium is both
hot and quite dense in these cases, the radio emission is very strong.
This agrees with our previous results from Section \ref{sec:phase},
where we found the bulk of the emission at late times to be in the
hot, dense phase of the gas.

\dofig{
\begin{figure*}[htpb]
  \centering
 \includegraphics[width=1.0\textwidth]{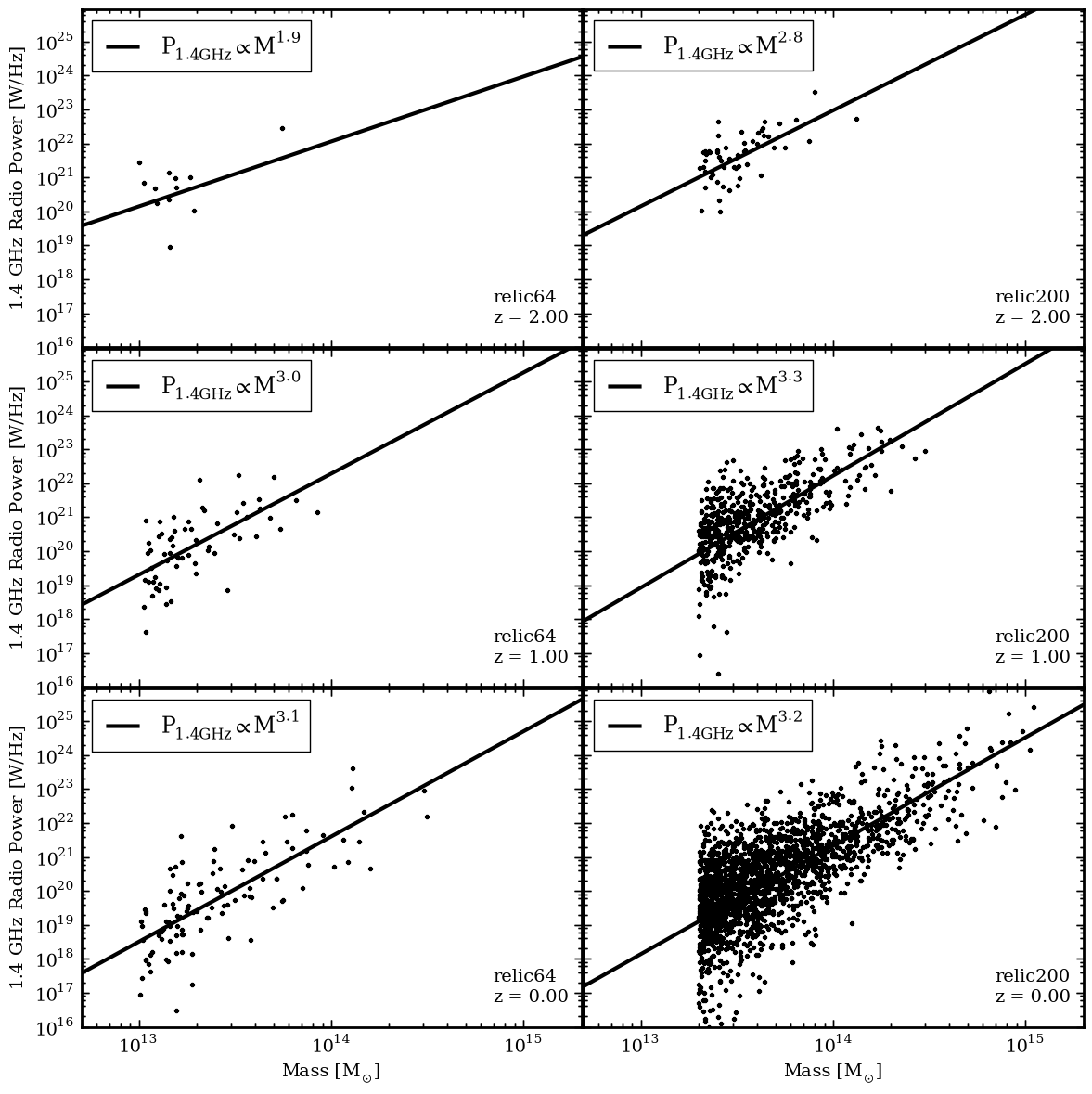}
 \caption{1.4 GHz Radio Luminosity-Mass Relationship of halos in
   \textit{relic64} (left) and \textit{relic200} (right) for redshifts
   2 (top), 1 (middle), and 0 (bottom). Shown in black are each
   individual halo. The lines represents a best power-law fit to all
   halos with $M_{vir}>10^{13}M_{\odot}$ and $2\times10^{13}M_\odot$
   for the $relic64$ and $relic200$ using a least-squares fitting
   routine.}
  \label{fig:lum_mass_64}
\end{figure*}
}

\subsection{Radio Power - Mass Relationship}\label{sec:radio_mass}

From the projections of individual halos in Figure \ref{fig:halos}, we
can see that there is a general trend for the more massive halos to
have higher radio emission (note masses decrease from the top left to
bottom right). We now want to quantify this scaling relationship by
studying the radio luminosity-mass relationship for the halos in our
simulations. We begin with the earlier list of halos and use the
virial quantities of each halo. For each halo, we use their total mass
and 1.4 GHz radio power (integrated out to $r_{200}$) to populate
Figure \ref{fig:lum_mass_64}. Second, for the distribution of halos,
we now determine the linear-least squares fit to
$\mathrm{log}(P_{1.4GHz}) = \mathrm{A} \mathrm{log}(M_{200}) +
\mathrm{B}$ for all halos with $M_{200} > 10^{13}M_\odot$ and
$2\times10^{13}M_\odot$ for the $relic64$ and $relic200$ simulations,
respectively. We choose to only fit halos above this minimum mass
because at smaller scales additional physics such as cooling not
included in our simulations would possibly strongly affect the
emission.  Additionally, we do not capture small mass halos that are
likely moving through these small clusters possibly creating a large
fraction of the total radio emission.  Because our simulation data
does not have a measurable uncertainty for a given radio power, we
have to use an alternate method of determining the error estimates of
our parameters.  We first find the best fit parameters using a uniform
weighting.  By calculating the residuals for each point from this
best-fit relation, we estimate the uniform error for each point as the
standard deviation of this residual.  We then fit the data again using
this error to obtain the uncertainty estimates in each parameter. The
values of these parameters are shown in Table \ref{tab:fitpars}.

\dofig{
  \begin{figure*}[htp]
    \centering
   \includegraphics[width=1.0\textwidth]{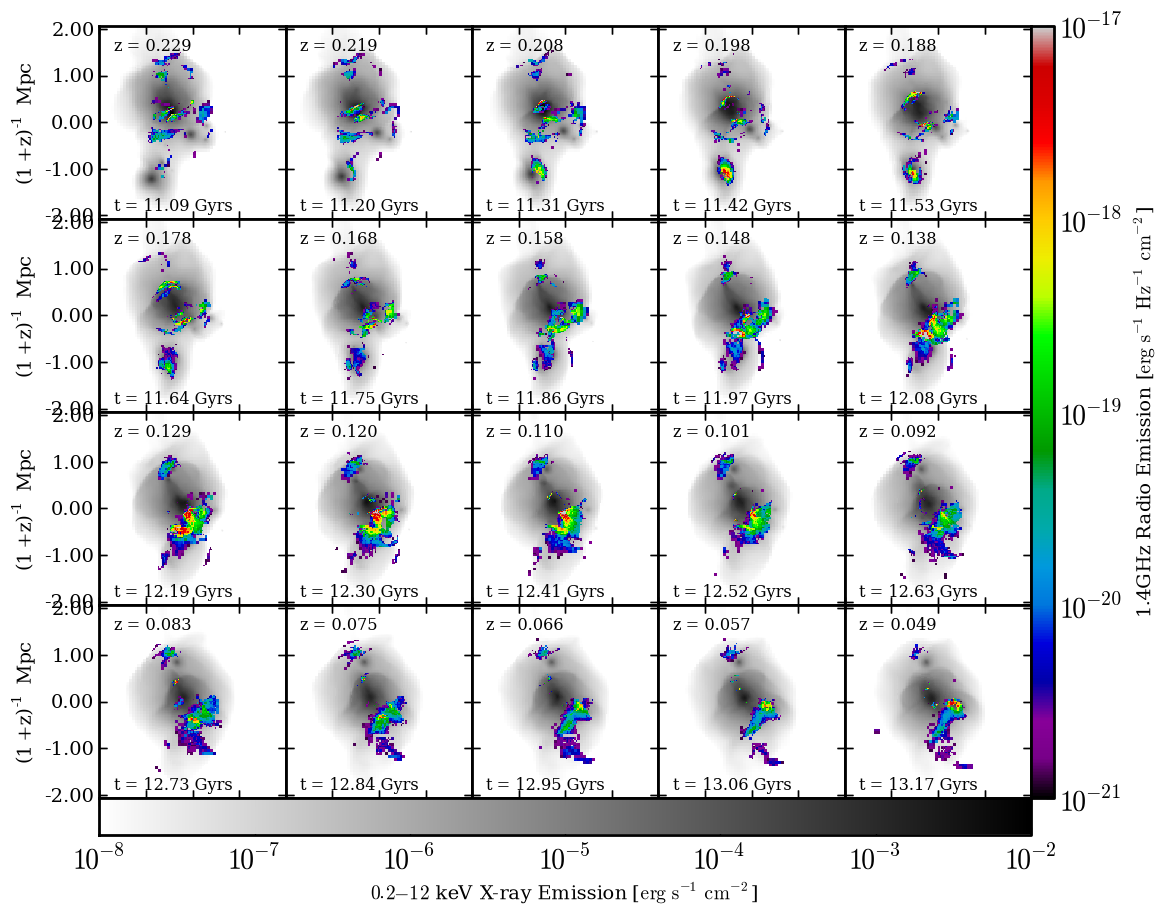}
    \caption{1.4 GHz Radio emission overlaid on 0.2-12 keV X-ray emission for 20 snapshots during merger activity of the largest cluster in the \textit{relic64} simulation from z=0.23 to z=0.05, a time span of 2 .08 billion years. Length units are comoving.}
    \label{fig:merger}
  \end{figure*}

}
\dofig{
  \begin{figure}[htpb]
    \centering
   \includegraphics[width=0.45\textwidth]{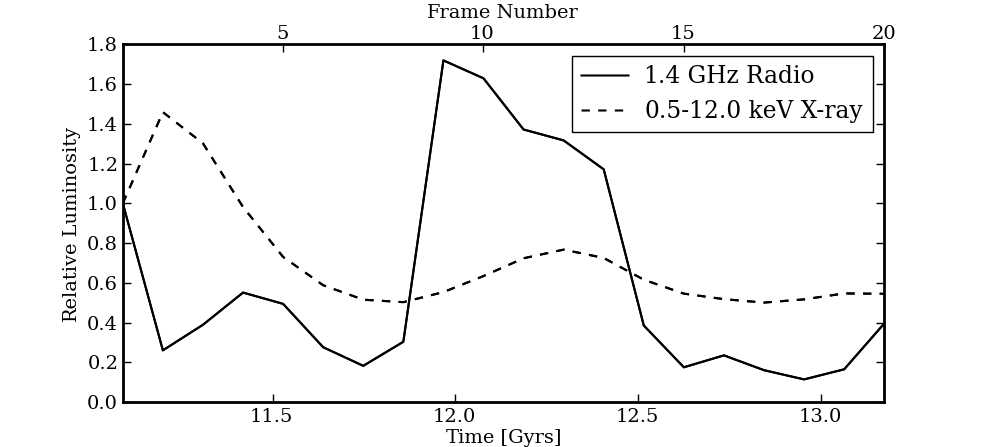}
    \caption{Time evolution of 1.4 GHz radio and 0.2-12 keV X-ray luminosities of the merger shown in Figure \ref{fig:merger}, normalized to the luminosities at $z=0.229$.  The first frames are a result of a prior merger in its last stages.}
    \label{fig:time-evolution}
  \end{figure}
}

Perhaps the most interesting result from Figure \ref{fig:lum_mass_64} is
the normalization as a function of redshift.  At $z=0$, a
$10^{14}M_\odot$ mass halo emits $\sim10^{20} erg/s/Hz$, while a
similar mass halo at $z=2$ emits $\sim10^{22} erg/s/Hz$.  This
amplifies our hypothesis that the merger state of the halo is very
important.  The $10^{14}M_\odot$ halo at $z=2$ is one of the most
massive objects at that time and has likely recently formed, whereas
the same mass halo at $z=0$ is a fairly common object that was likely
formed some time ago.  Additionally, the probability of a merger with
a $1:1$ mass ratio is very low for the largest halos at a given
time. For both simulations at multiple redshifts, the radio power is
correlated with the mass of the halo. This dependence is expected
since the radio emission is a function of the temperature, density and
magnetic field strength of the halo, all of which scale positively
with the mass.

The second result of these radio luminosity-mass relationships is the
large scatter around the best fit. We see that there can be scatter of
up to $2-3$ orders of magnitude for a given mass cluster. This
suggests that while the radio emission is correlated with mass, the
merger state of the halo plays a major role in determining the radio
power.  As can be seen from the projections of these halos in Figure \ref{fig:halos}, the
most radio luminous objects have very disturbed morphology and are
undergoing major mergers.

In Figure \ref{fig:merger}, we show the evolution of the most massive
cluster in \textit{relic64}, and track the total radio and X-ray
luminosity as a function of redshift.  The X-ray emission is
calculated using outputs of the Cloudy code \citep{Ferland:1998aa}
where we have adapted the method of \citet{Smith:2008aa} for radiative
cooling to calculate frequency-dependent emission. This yields an
X-ray emission for given temperature and density of the gas, and is
shown in gray.  The radio emission, shown in color, has been masked
such that all values below $10^{-21} erg /s~Hz^{-1} cm^{-2}$ are
transparent, allowing for a view of the X-ray data and masking out
radio features that are too faint to be observed.  To help follow the
evolution of the total radio and X-ray luminosities from the cluster,
we plot their relative luminosities with respect to their values at
$z=0.23$ in Figure \ref{fig:time-evolution}.   

As we follow the evolution from $z=0.23$, we see the evolution of a
major merger where the two cores pass through each other at $z=0.22$.
The smaller halo is moving from the center towards the upper-right.
As the shockwave builds through $z=0.17$, the radio emission closely
follows the X-ray brightness jump, as we would expect. By $z=0.16$,
the radio emission from the initial shock has decreased dramatically.
While the initial shock has disappeared in the radio, a secondary
shock has been created that moves in from the hot ICM into the wake
the merger left behind from the middle towards the lower-right.  By
$z=0.15$ (frame 9), this is the most luminous feature in the radio.
At this time, the image is 7-8 times brighter in the radio than it was
at $z=0.16$ (frame 8), illustrating how strongly radio emission
depends on the merger state of the cluster.  As the halo evolves
further, additional smaller objects fall into the ICM, but don't get
nearly as bright as major merger.  By $z=0.1$ (frame 14), the
integrated radio luminosity has dropped back to pre-merger levels.

During this merger, the X-ray luminosity also increases, but the total X-ray
emission only increases by $~50\%$, which again illustrates the difference
in the X-ray and radio emission mechanisms. A detailed analysis of
cluster evolution and merger state is reserved for a later study.

\dofig{
\begin{figure*}[htp]
  \centering
\includegraphics[width=1.0\textwidth]{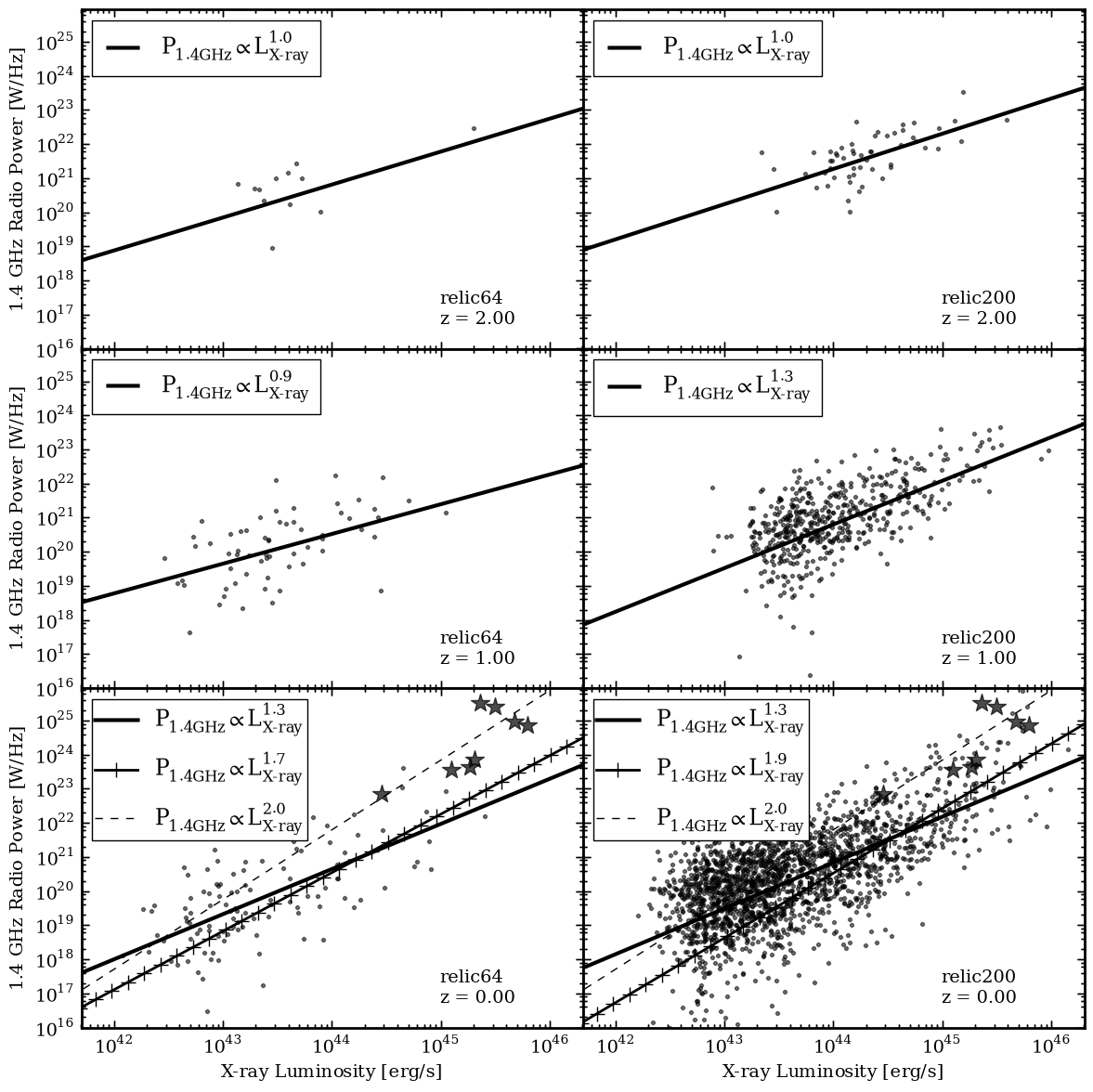}
\caption{$P_{1.4GHz}-L_X$ relationship for halos in \textit{relic64}
  (left) and \textit{relic200} (right) for $z=2,1,0$ from top to
  bottom.  Both radio and X-ray emissivity are integrated out to the
  virial radius for each halo.  A best fit line is found for halos
  with $M_{vir}>10^{13}M_{\odot}$ and $2\times10^{13}M_\odot$ for the
  $relic64$ and $relic200$ using a least-squares fitting routine.  For
  $z=0$, we show fits to our data using a minimum mass (solid) as above and
  minimum X-ray luminosity of $10^{44} erg/s$ (solid + crosses).  Also shown are
  observational data (stars) from \citet{Feretti:2002aa} along with a
  best fit (dashed).  }\label{fig:radio_xray}
\end{figure*}
}

\subsection{Radio Power-X-ray Relationship}\label{sec:radio_xray}

\dofig{
\begin{table*}[htp]
  \centering
  \begin{tabular}{ c | c | c | c | c | c | c | c | c | c }
   \multicolumn{2}{c}{Simulation} & \multicolumn{4}{|c|}{\textit{relic64}} & \multicolumn{4}{|c}{\textit{relic200}} \\ 
    \hline
    $z$ & Var & A & $\sigma_A$ & B & $\sigma_B $&  A  & $\sigma_A$ & B & $\sigma_B$ \\ 
    \hline
    \hline
    $z=0$ &Mass&  3.1 & 0.1 & 18.51 & 0.07 & 3.19 & 0.04 & 18.14 & 0.02 \\
    $z=1$ &Mass&  3.0 & 0.2 & 19.38 & 0.08 & 3.30 & 0.08 & 18.93 & 0.04 \\
    $z=2$ &Mass&  1.9 & 0.4 & 20.1 & 0.1    & 2.8 & 0.2 & 20.2 & 0.1  \\
    \hline
    $z=0$ &X-ray& 1.32 & 0.08 & 19.32 & 0.06 & 1.34 & 0.02 & 19.49 & 0.02 \\
    $z=1$ &X-ray&  0.9 & 0.2 & 19.6 & 0.2         & 1.28 & 0.04 & 19.52 & 0.04  \\
    $z=2$ &X-ray& 1.0 & 0.3 & 19.8 & 0.2         & 1.0 & 0.1 & 20.2 & 0.1  \\
 \end{tabular}
  \caption{Best fit parameters for radio power scaling relationship with halo mass and 0.2-12 keV X-ray emission.  Fitting functions are $10^{B} (M_{200}/10^{13}M_\odot)^A$ and $10^{B} (L_x/10^{43}erg/s)^A$ for Mass and 0.5-12 keV X-ray luminosity, respectively.}
  \label{tab:fitpars}
\end{table*}
}

While the X-ray and radio emission may not be coincident in projection,
we expect to see a correlation between the total X-ray and radio
luminosity since they both sample hot, dense gas. We start our
analysis from the results of our radial profiles and examine the $0.2-12keV$ X-ray
and $1.4~GHz$ radio emission within $r_{200}$ for \textit{relic64}
and \textit{relic200} in Figure \ref{fig:radio_xray}. 

We again use a method of linear-least squares regression described
above to obtain a scaling relationship between the X-ray and radio
luminosity, the results of which are shown in Table \ref{tab:fitpars}.
Again we see that while there is a clear trend with X-ray luminosity,
large scatter in the individual clusters can dominate the
relationship. This scatter likely comes from two sources. First, a
cluster that is relatively relaxed will have significant X-ray
emission, whereas the lack of shocks in such a scenario will
necessarily lead to zero radio emission in this model. Second, the
fractional increase in X-ray luminosity across a shock front is much
less than the radio emission because the X-ray primarily depends on
the density of the gas at cluster temperatures, whereas the radio
scales with density \textit{and} temperature.

As a function of redshift, the scaling relationship between radio
power and X-ray luminosity evolves much like the radio power-mass
relationship.  Objects with the same X-ray luminosity at early times
are more likely to have much higher radio power than their
low-redshift counterparts.  Additionally, the strength of the
relationship increases at low redshift significantly due to the
stronger correlation with larger mass halos.

Even though our constraints on the fit parameters seem quite small,
one can argue the due to the large scatter there should be a broad
range of values capable of producing ``acceptable'' fits.  To give a
basic understanding of how our fit parameters can vary, for our two
simulations at $z=0$, we have shown the results of fitting using two
cuts on our underlying data.  The solid line is the result of fitting
the points using all halos with amass greater than $10^{13}M_\odot$
and $2\times10^{13}M_\odot$ for the $relic64$ and $relic200$
simulations, respectively.  The solid line with hash marks shows the
result of only fitting points with $L_X>10^{44}erg/s$.  As one can
see, the slope varies quite dramatically.  Therefore when comparing to
observational constraints, the selection function of the
observed/simulated clusters is very important.

We can compare our derived scaling relationships with observational
estimates from known radio relics.  \citet{Feretti:2002aa} found X-ray
luminosities and $1.4GHz$ radio power for 9 Abell clusters.  If we fit
their data using our same least-squares regression technique, we
obtain $P_{1.4GHz} \propto L_X^{2.0\pm0.5}$, agreeing quite well
within the uncertainties in our $z=0$ simulation data.  However, the
normalization for the real relics are much higher than our simulated
relics.  We explicitly plot these clusters on Figure
\ref{fig:radio_xray}.  A very important point from this is that the
observed relics land on the high end of both the X-ray luminosity and
radio power.  This demonstrates the selection effects coming in to
play, as we have only observed the brightest objects as of yet.  This
suggests that deeper observations of radio-quiet clusters should lead
to the discovery of low power radio relics.  As another constraint,
\citet{Cassano:2006aa} study this relationship for giant radio halos
and find $P_{1.4GHz} \propto L_X^{1.74 \pm 0.21}$.  While these giant
radio halos are thought to be from turbulent re-acceleration of
electrons, their origin is likely linked to the same driving forces
(i.e. mergers) as the radio relics.  We note here that our X-ray
emission is likely underestimated due to our lack of radiative physics
in our simulations.  However, properly modeling galaxy formation,
metal pollution, cooling, and thermal feedback is beyond the scope of
this study.

\dofig{
\begin{figure*}[htp]
\centering
\includegraphics[width=0.49\textwidth]{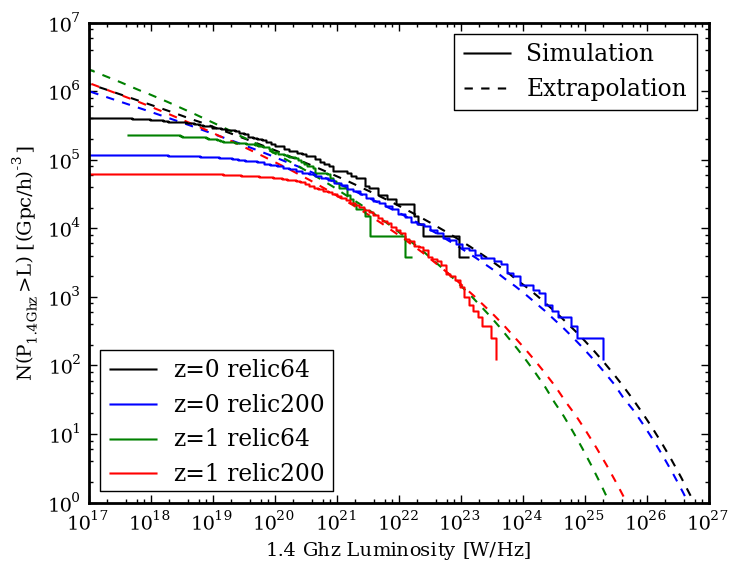}
\includegraphics[width=0.49\textwidth]{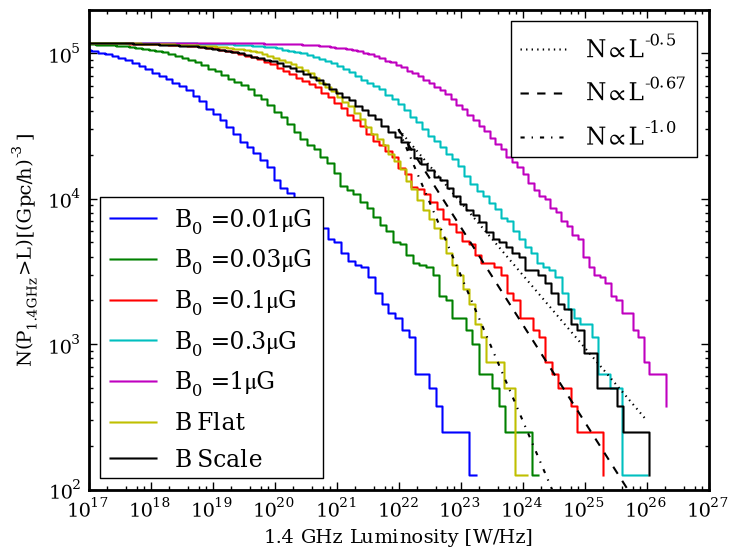}
\caption{(Left) 1.4 GHz Radio Luminosity Function for clusters in
  \textit{relic64} and \textit{relic200} for $z=0,1$, in units of
  inverse comoving $(\mathrm{Gpc}/h)^3$. Shown in dashed lines are
  extrapolations using the \citet{Warren:2006aa} mass function and our
  $\mathrm{P_{1.4GHz} - Mass}$ scaling from Table
  \ref{tab:fitpars}. (Right) The Radio Luminosity Function for
  clusters in \textit{relic64} as a function of magnetic field
  model.}\label{fig:lum_func}
\end{figure*}
}

\subsection{Luminosity Function}\label{sec:luminosity_function}

Now that we have explored how the radio emission varies as a function
of mass and X-ray luminosity, we ask the question: ``How many radio
relics do we expect at a given luminosity in the Universe?'' Here we
attempt to answer this question by constructing an
observationally-motivated radio luminosity function. This is done by
calculating the cumulative number of objects brighter than a given
luminosity.  We have done so in Figure \ref{fig:lum_func}, and
normalized the count rates by $ h^3(1+z)^3~Gpc^{-3}$.

As we expect from our radio luminosity-mass relationship, the overall
shape of the luminosity function is similar to the cluster mass
function presented in Section \ref{sec:radio_mass}. At first glance
this luminosity function is not very encouraging for observational
studies because even our most luminous objects are difficult or
impossible to capture with current radio telescopes. However, this is
primarily a result of the mass range in our current simulations. The
known radio relics are associated with massive clusters with
$M>10^{15}M_\odot$, and our largest clusters in \textit{relic200} only begin
to reach the $10^{15}M_\odot$ mark. Obtaining spatial resolution
needed to capture the relics in a volume large enough to capture these
very rare clusters is computationally difficult. We address this by
combining best estimates of the halo mass function with our radio
luminosity-mass relationship.

We begin by calculating the best fits for the radio luminosity-mass
relationships for both of our simulations.  Next, we take fits from
\citet{Warren:2006aa} for the mass function at redshift 0 and 1.  We
then convert the mass in the mass function to the expected radio
luminosity from our fits in Table \ref{tab:fitpars}.  The results of
this fitting are shown in the left panel of Figure \ref{fig:lum_func}.
Because of the scatter in the radio-mass luminosity function, we are
able to place rough lower and upper limits on the luminosity function.
This scatter will be constrained by future simulations that cover a
larger mass scale of galaxy clusters.

As we have done for our other results in Appendix A, we varied the
magnetic field model to examine its effects on the luminosity
function.  The first parameter we changed is the normalization of the
magnetic field, $\mathrm{B_0}$.  In Figure \ref{fig:lum_func}, we show
$\mathrm{B_0=\{0.01,0.03,0.1,0.3,1.0\}\mu G}$.  Since the emitted
power is roughly proportional to $B^{5/2}$ (see Section 3.1), as we
increase $\mathrm{B_0}$ the luminosity function shifts quite
dramatically to larger luminosities.  At low values of $B_0$ the
increase is close to the expected increase of $B^{5/2}$, while at
higher values $B$ approaches $B_{CMB}$, reducing the effect of the
increased local field strength.  The second variation was in the
scaling of the magnetic field with respect to the electron density.
The line labeled ``B-Flat'' corresponds to $B=B_0$, whereas
``B-Scale'' denotes $B \propto B_0 n_e$.  In both cases we set
$B_0=0.1\mu G$.  With a uniform magnetic field, we see that the number
of high-luminosity objects decreases dramatically, while the number of
low luminosity objects increases slightly.  This is understandable
given that the highest luminosity objects come from the most massive
clusters, which have the highest densities.  In this case, the density
doesn't correspond to higher magnetic fields, and the radio luminosity
is diminished with respect to the adiabatic scaling.  Similarly, in
the ``B-Scale'' case, the magnetic field strength is even higher in
the dense parts of the largest clusters, leading to a shallower slope
in the luminosity function.  By comparison, \citet{Hoeft:2008aa} also
found a slope of $-2/3$ using the same model as the $B_0=0.1\mu G$
line in \ref{fig:lum_func}, adding verification to both results.  

To determine the number of clusters for a given survey area and
redshift depth, we integrate the cosmological volume out to $z=0.5$ for
a given survey area $d\Omega$,
\begin{eqnarray}\label{eq:volume}
  \frac{dV}{dz d\Omega}(z) = \frac{c}{H_0} \frac{(1+z)^2D_A^2}{E(z)}
\end{eqnarray}
with
\begin{eqnarray}
  E^2(z) = \Omega_{m,0}(1+z)^3 + \Omega_\Lambda,
\end{eqnarray}
where $D_A$ is the angular diameter distance.  The result of this is
that an all-sky survey out to $z=0.5$ covers $26.1
~(\mathrm{Gpc}/h)^3$.  In combination with our estimates from the
$relic200$ simulation in Figure \ref{fig:lum_func}, we expect to find
180 (conservative) to 1000 (optimistic) clusters with a total radio
luminosity of $10^{25}W/Hz$ within this cosmological volume.

We also see from Figure \ref{fig:lum_func} that the luminosity
function of halos increases from $z=1$ to $z=0$.  However, if plotted
using proper volumes, the factor of 8 brings the two luminosity
functions much closer together.  Therefore the proper number density
of radio relics seems to be fairly constant through cosmic time. This
is an unexpected result, and encouraging for moderate redshift studies
of radio relics. We note here that the frequency at
which telescopes receive this synchrotron emission changes as a
function of the emitter's redshift. Therefore when deriving the radio
luminosity of an object at redshift $z$, we use $\nu =
1.4~\mathrm{GHz}~(1+z)$. Because of this, the emitted power is
actually decreased since $P_{1.4GHz}\propto \nu^{-s/2}$ where
$s\approx2$ for strong shocks. The power emitted in the cluster's
frame is therefore substantially larger than what is shown in Figure
\ref{fig:lum_func}. The similar luminosity function is therefore a
product of the increased merger and accretion activity at higher
redshift compared to that at $z=0$.

\section{Discussion}\label{sec:discussion}

\subsection{Comparison To Previous Work}\label{sec:append_Flux}

In order to compare our results to previous shock studies, we have
calculated the kinetic energy flux through shocks, as shown in figure
\ref{fig:energy_radio_flux}.  The left two panels show the kinetic
energy flux as a function of Mach number in the $relic64$ and
$relic200$ simulations, respectively.  The right panels instead show
the radio emission as a function of Mach number.  While the black
lines denote all temperatures, we also show the breakdown in terms of
the pre-shock temperature.  The kinetic energy flux results here can
be directly compared to Figure 6 of \citet{Ryu:2003aa}, Figure 10 of
\citet{Skillman:2008aa}, Figure 11 of \citet{Vazza:2009aa}, and can
also be compared after unit conversions to Figure 6 of
\citet{Pfrommer:2006aa}.  Even though these simulations all vary in
size and adopted cosmological parameters, the similarities in the
kinetic energy flux processed by shocks is quite strong.  This
suggests that the underlying shock characteristics are quite well
understood even if the particular radio emission models vary.

\dofig{
\begin{figure*}[htp]
  \centering
 \includegraphics[width=\textwidth]{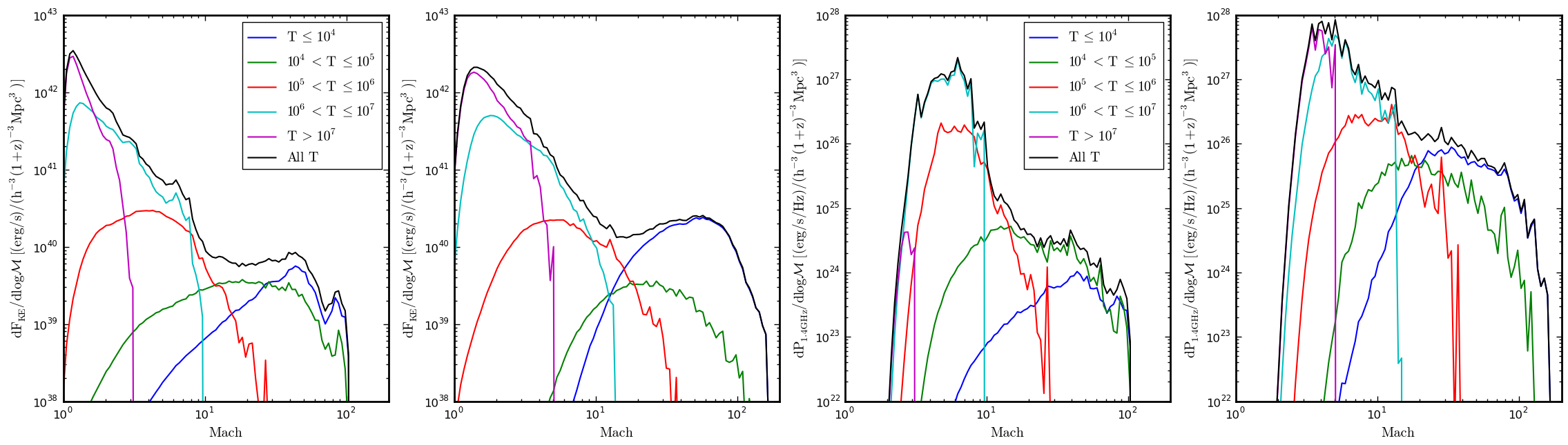}
\caption{Left Panels: $relic64$ and $relic200$ kinetic energy flux processed by shocks at $\mathrm{z=0}$.  Right Panels: 1.4 GHz radio emission.}
  \label{fig:energy_radio_flux}
\end{figure*}
}

In the right panels of \ref{fig:energy_radio_flux}, we can see that
there is a larger difference between the two simulations presented
here in terms of the radio emission.  This is likely due to the
varying mass scales present in the simulations.  We have also
subsampled the $relic200$ simulation into random $(\mathrm{64
  Mpc/h})^3$ domains and found that a major contribution is confined
to one of these subdomains.  

One of the earliest studies of shocks in a cosmological context is
found in \citet{Miniati:2000aa}, where the authors found similar
shock structure and kinetic energy flux trends as is seen in this
study, though in a unigrid context.  They, too, found that intermediate
Mach number shocks are responsible for processing the majority of
kinetic energy.  In a pioneering work, \citet{Miniati:2001ab}
studied the injection and evolution of cosmic ray electrons.  Using a
framework to follow the cosmic ray distribution, they presented a
radio power - core temperature relationship that shows strong
similarity to what we have found with respect to cluster mass and
X-ray luminosity, including a larger amount of scatter from
cluster-to-cluster.  Even though the resolution was modest compared to
studies here, many of the primary characteristics of the radio
emission are similar.  

Much of our work presented here can be compared with that of
\citet{Hoeft:2008aa}.  We use the same radio emission model, but
instead apply it to AMR simulations as opposed to smoothed particle
hydrodynamic simulations. In particular, we can compare our radio
relic luminosity function in Figure \ref{fig:lum_func} to their Figure
9.  After accounting for the different normalization, we find that we
have more objects at $\sim 10^{25} W/Hz$.  However, this result is from a
small number of objects in our simulations and therefore future
simulations with a larger sample of galaxy clusters are needed.

In \citet{Pfrommer:2008aa}, the authors studied the acceleration and
emission properties of cosmic ray electrons and protons in a smoothed
particle hydrodynamics setting which focused on a set of high resolution
galaxy clusters.  Many of the same characteristics of galaxy cluster
radio relic emission that we found in this study are consistent with
their results.  The morphology of the radio relic emission is very
similar to our results, though since they follow the electron
population through time the emission is more diffuse compared to our
simulated clusters.  In another paper in the same series,
\citet{Pfrommer:2008ab} study the scaling relationship between
the radio synchrotron, gamma-ray, and inverse Compton emission from the
same set of galaxy clusters.  Their results when fitting the scaling
relationship between radio synchrotron emission and cluster mass give a
significantly shallower slope of $1-1.5$.  However, due to the small
number of clusters in their study, it is difficult to tell if there
is a meaningful difference between their results and the ones presented
here.  In future work it would be useful to run a series of high resolution AMR
simulations using our methods to compare to their results.  

\subsection{Implications For Future Surveys}\label{sec:implications}

Our study has shown that nearly every cluster has radio emission and
displays signs of radio relics at some stage in their evolution.  When
and where this radio emission occurs, however, is very sensitive to
the merger and evolutionary state of the cluster.  Current studies of
radio relics have been confined to pointed observations of nearby,
massive clusters, often based on strong X-ray emission.  While this
observational strategy does conform to our general results found in
mass and X-ray scaling relationships, we have determined that not all
X-ray luminous or massive clusters have significant relic emission.
It is instead heavily biased towards disturbed, merging clusters.
Because the surface brightness of these relics is low due to their
extended nature, large surveys with near-future telescopes are
unlikely to yield serendipitous discoveries of cluster radio emission.
Instead, the focus should be on deep, multiwavelength, large
field-of-view observations with sensitivity to extended diffuse radio
emission of disturbed X-ray clusters.

Additionally, studies must include regions away from the peak X-ray
emission.  As was seen in Section
\ref{sec:individual_object_properties}, radio emission from shocks is
generally brightest at the edges of clusters, surrounding the X-ray
emission.  This prescribes a fairly difficult observational roadmap,
but the potential benefits include studying fundamental plasma physics
phenomena such as in-situ shock electron acceleration and magnetic
field structure.  By combining statistical and morphological studies
of these objects, we can readily compare them with the high-resolution
hydrodynamical simulations presented here.

Our luminosity function, derived from the \citet{Warren:2006aa} mass
function, suggests what can be expected in future observational studies.
With surface brightness sensitivity improvements of a factor of 10, we can expect to
see an increase of a factor of 10-100 in the number of clusters with
radio relics.  Alternatively, an all-sky survey out to $z=0.5$ should
result in the discovery of $\sim200$ clusters above a luminosity of $10^{25} W/Hz$.

\subsection{Limitations of the Models}\label{sec:limitations}

There are several limitations to this study.  First, we have not
performed self-consistent MHD simulations, and instead adopted a
scaling relationship between the post-shock density and magnetic
field.  This leads to the absence of any magnetic field configurations
imprinting their structure on the radio relics.  We plan to
incorporate MHD simulations in future work.  However, since $B \sim
B_{CMB}$ in the cluster environments we study here, the magnitude of
our radio emission should not change dramatically. We are also not
presenting a self consistent view of magnetic field generation,
evolution, and non-linear interactions with shocks or particle
acceleration, all of which are poorly understood in a cosmological
context.  

Second, due to our relatively small box sizes, we have not
captured the largest objects in the universe, which are likely to
produce the brightest radio signatures.  To account for this, we have
provided an estimate using an analytic mass function combined with
mass-luminosity scaling relationships.  However, self-consistently
capturing these very massive clusters is important, and will be
explored in future work. Third, we also only follow recently
accelerated electrons and ignore the aging of electron populations or
re-accelerated electrons, which is likely to be important for radio
halos and steep-spectrum objects.  A more realistic model would follow
the electron population ``on-the-fly'' and modify the acceleration
efficiency as a function of pre-existing electron populations.  Given
that we assume radio emission only comes from electrons that have just
been accelerated, this means that the radio relic luminosities shown
are lower limits.

While our spatial and mass resolutions are quite good in the $relic64$
simulation, it is likely that increasing the resolution would have an
affect on our results.  \citet{Skillman:2008aa} found that in terms of
the kinetic energy processed by shocks, a peak spatial resolution of
$3.9 kpc/h$ and a mass resolution of roughly $10^9 M_\odot$ was
approaching a converged result, though perfect convergence was not
seen.  In $relic64$ our spatial resolution matches this value, while
we are a factor of two above this mass resolution.  For $relic200$, we
are likely not capturing all of the kinetic energy flux in low Mach
number shocks, which would suggest a higher spatial and mass
resolution simulation would lead to somewhat higher radio luminosity emission.
This will likely always be the case since any increase in mass
resolution will lead to a greater sampling of the mass function,
allowing one to follow the merger assembly of galaxy clusters more
accurately.  This should increase the frequency of merger shocks,
increasing overal radio emission.

Finally, the electron acceleration efficiency in Equation \ref{eq:2}
is poorly constrained at present.  With additional radio observations,
particularly using next generation low frequency radio telescopes,
along with new PIC simulations, we may be able to
calibrate $\xi_e$ to more accurate values.  This could be important
in scaling the radio luminosity function in Figure \ref{fig:lum_func}
and estimating the number of radio relics expected in clusters and/or
sky surveys.

\section{Conclusions and Future Directions}\label{sec:conclusions}

We have carried out high resolution AMR cosmological simulations using
our accurate shock finding algorithm with a radio emission model for
shock-accelerated electrons to examine the properties of radio relics in
galaxy clusters.  From this model, our main results are:

\begin{itemize}
\renewcommand{\labelitemi}{$\bullet$}

\item We have produced synthetic radio maps of the large scale structure
  and cluster environments, showing the variety of radio relic
  morphologies and locations.

\item Through the use of 2D phase diagrams, we have found that while
  there is radio emission from both merger (internal) and accretion
  shocks, the emission from the hot, dense intracluster medium
  associated with the merger shocks dominate the total emission.  This
  balance is redshift-dependent, with accretion shocks being more
  important at high redshift.

\item We have generated scaling relationships using over 2000
  simulated halos that give insight to how radio emission scales with
  mass and X-ray luminosity.  These relationships evolve with redshift
  and there is a large scatter for individual halos as a result of
  merger state.

\item By studying the time evolution of a cluster undergoing a merger,
  we find that the radio emission is highly dependent on the merger
  state, varying on time scales of a few hundred million years.

\item We have produced a synthetic radio luminosity function that gives
  observational predictions for the number of clusters with radio
  relics. This can be used to compare to future observed cluster
  luminosity functions and as a test of synchrotron emission models.
\end{itemize}

In future studies we plan to examine the merger history and morphology
of these objects in greater detail.  The redshift
evolution of individual clusters is likely to be heavily correlated
with merger state, and a statistical study of this relationship is
vital to future observational studies.  Finally, in order to correctly
model these radio relics, we need to self-consistently follow the
electron population, taking into account effects of particle aging and
re-acceleration.  Future studies will also examine larger cosmological
volumes and implement techniques such as light cones.

\acknowledgements{ S.W.S would like to thank Matthias Hoeft and Marcus
  Br\"{u}ggen for making their radio emission model available.  S.W.S
  and B.W.O. thank Megan Donahue and Mark Voit for useful
  conversations and comments.  We also thank an anonymous referee for
  very helpful comments and suggestions.  Computing time was provided
  by NRAC allocations TG-AST090040 and TG-AST090095. S.W.S.,
  E.J.H. and J.O.B. have been supported in part by a grant from the US
  National Science Foundation (AST-0807215).  S.W.S. has been
  supported by a DOE Computational Science Graduate Fellowship under
  grant number DE- FG02-97ER25308.  B.W.O. has been supported in part
  by a grant from the NASA ATFP program (NNX09AD80G). E.J.H. also
  acknowledges support from NSF AAPF AST 07-02923.  B.D.S has been
  supported by NASA grant NNZ07-AG77G and NSF AST0707474.
  Computations described in this work were performed using the
  \textit{Enzo} code developed by the Laboratory for Computational
  Astrophysics at the University of California in San Diego
  (http://lca.ucsd.edu) and by a community of independent developers
  from numerous other institutions.  The \texttt{yt} analysis toolkit
  was developed primarily by Matthew Turk with contributions from many
  other developers, to whom we are very grateful.  The LUNAR
  Consortion (http://lunar.colorado.edu), headquartered at the
  University of Colorado, is funded by the NASA Lunar Science
  Institute (via cooperative Agreement NNA09DB30A). The authors
  acknowledge the Texas Advanced Computing Center (TACC) at The
  University of Texas at Austin for providing HPC resources that have
  contributed to the research results reported within this paper. URL:
  http://www.tacc.utexas.edu.  Some computations were also performed
  on Kraken (a Cray XT5) at the National Institute for Computational
  Sciences (http://www.nics.tennessee.edu/).}

\appendix

\section{Varying Magnetic Field Models}\label{sec:append_emission}

Here we briefly examine the impact of an alternate magnetic field
model.  We will concentrate on two of our results, and leave further
in-depth analysis to future work.  In both cases, we present 6
alternate models.  The first four concentrate on changing the reference
magnetic field value, $\mathrm{B_0=\{0.01,0.03,0.3,1.0\}\mu G}$.  The
second two models test the dependence of the magnetic field with
respect to the electron number density.  For these two cases, we chose
simple parameterizations that simply give a feel for how this
relationship affects our results.  In one case, we keep the magnetic
field constant throughout the domain at $B=B_0=0.1\mathrm{\mu G}$.  In
the second case, we scale the magnetic field proportionally with the
number density $B = B_0 (n_e/n_{e,0})$, with $\mathrm{B_0=0.1\mu G}$
and $\mathrm{n_{e,0} = 10^{-4} cm^{-3}}$.  

First we examined the fundamental impact on the phase of the gas
responsible for the radio emission.  In Figure
\ref{fig:bfield-phase} where we show the relative emission to our
fiducial parameters, the primary effect of lowering $B_0$ is to
decrease the total emission as roughly $B^{5/2}$, as expected.  If we
instead increase $B_0$, because it starts to become comparable to
$B_{CMB}$, the emission enhancement is preferably increased in the
high temperature, low-Mach number regime corresponding to merger
shocks.  In the case where we have a flat magnetic field, the emission
in the high temperature, high density, low Mach number regions is
greatly diminished.  However, the accretion shock emission is
increased by a factor of 100-1000.  If we let the magnetic field scale
proportionally to number density, the opposite is true.  Accretion
shocks have weaker radio emission, whereas the merger shocks are more
luminous.

If we instead examine the effect of a changing magnetic field model on
the radio luminosity-mass relationship, we again find a coherent
picture, as shown in Figure \ref{fig:bfield-halos}.  The key concept
is that as the magnetic field approaches $B_{CMB}$, the added emission
per unit increase is diminished by the second $B^2$ term in the
denominator of Eq. (\ref{eq:2}).  Therefore, for low values of $B_0$
when the magnetic field is much lower than $B_{CMB}$, the higher
magnetic field strengths in larger clusters has more of an effect,
steepening the radio luminosity-mass relationship.  On the other hand,
when $B_0$ is increased, the relative gains in magnetic field in large
clusters does not impact the emission as strongly, flattening the
relationship.  In changing the $B\propto n_{e}$ relationship, a flat
magnetic field removes the density bias between large and small
clusters, thereby flattening the scaling relationship.  A linear
scaling with number density increases this bias, steepening the
relationship.  

\dofig{
\begin{figure}[htp]
  \centering
\includegraphics[width=1.0\textwidth]{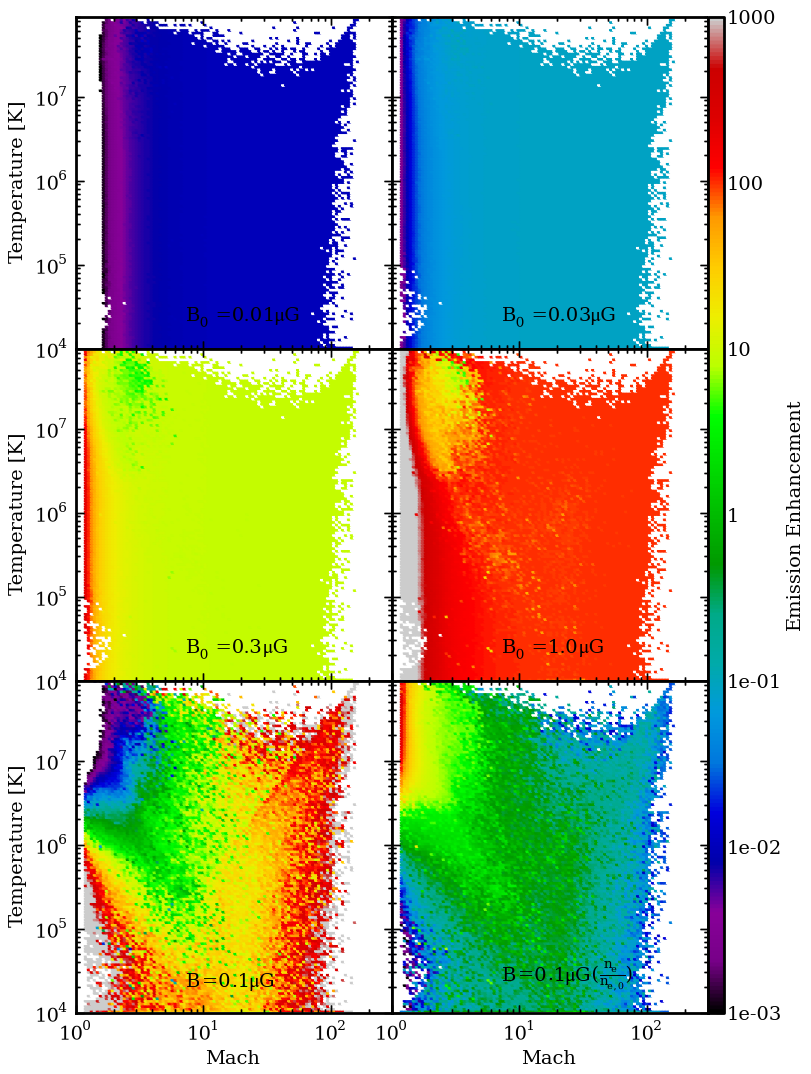}
\caption{Relative radio emission with respect to our fiducial magnetic field model for the $relic200$ simulation at $z=0$, shown in Figure \ref{fig:mach-temp-radio}.  In the upper four panels, the reference magnetic field parameter, $B_0$ is varied.  In the bottom left panel, the magnetic field strength is flat at $B=0.1\mu G$.  In the bottom right panel, the magnetic field strength scales linearly with electron number density.}
 \label{fig:bfield-phase}
\end{figure}
}
\dofig{
\begin{figure}[htp]
  \centering
 \includegraphics[width=1.0\textwidth]{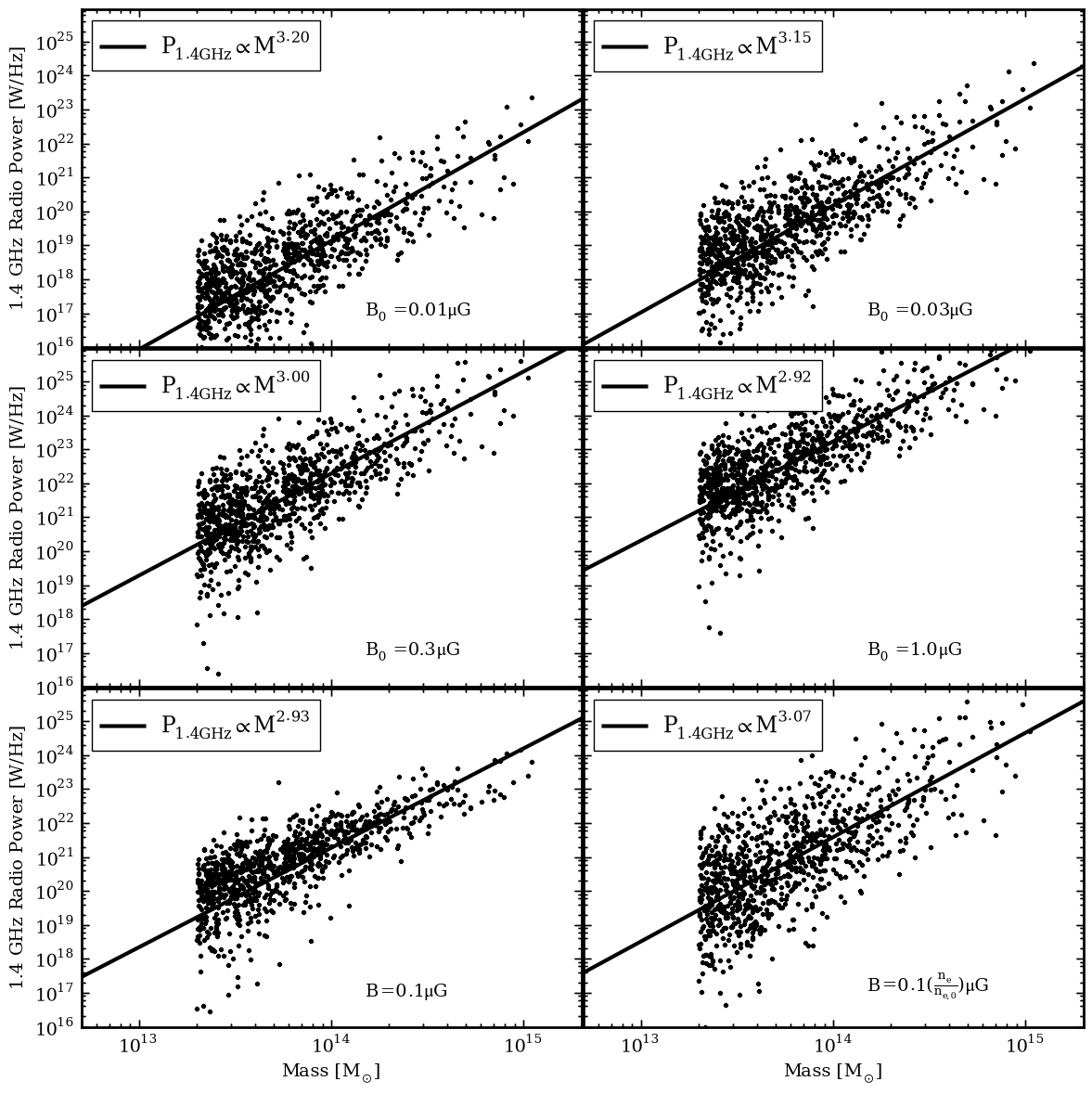}
\caption{As in Figure \ref{fig:lum_mass_64}, but for varying magnetic field model parameters.  In the upper four panels, the reference magnetic field parameter, $B_0$ is varied.  In the bottom left panel, the magnetic field strength is flat at $B=0.1\mu G$.  In the bottom right panel, the magnetic field strength scales linearly with electron number density.}
 \label{fig:bfield-halos}
\end{figure}
}

\vspace{5mm}
\bibliography{main}

\begin{thebibliography}{61}
\expandafter\ifx\csname natexlab\endcsname\relax\def\natexlab#1{#1}\fi

\bibitem[{{Aleksi{\'c}} {et~al.}(2010){Aleksi{\'c}}, {Antonelli}, {Antoranz},
  {Backes}, {Baixeras}, {Balestra}, {Barrio}, {Bastieri}, {Becerra
  Gonz{\'a}lez}, {Bednarek}, {Berdyugin}, {Berger}, {Bernardini}, {Biland},
  {Bock}, {Bonnoli}, {Bordas}, {Borla Tridon}, {Bosch-Ramon}, {Bose}, {Braun},
  {Bretz}, {Britzger}, {Camara}, {Carmona}, {Carosi}, {Colin}, {Commichau},
  {Contreras}, {Cortina}, {Costado}, {Covino}, {Dazzi}, {De Angelis}, {De Cea
  del Pozo}, {De los Reyes}, {De Lotto}, {De Maria}, {De Sabata}, {Delgado
  Mendez}, {Doert}, {Dom{\'{\i}}nguez}, {Dominis Prester}, {Dorner}, {Doro},
  {Elsaesser}, {Errando}, {Ferenc}, {Fonseca}, {Font}, {Galante},
  {Garc{\'{\i}}a L{\'o}pez}, {Garczarczyk}, {Gaug}, {Godinovic}, {Hadasch},
  {Herrero}, {Hildebrand}, {H{\"o}hne-M{\"o}nch}, {Hose}, {Hrupec}, {Hsu},
  {Jogler}, {Klepser}, {Kr{\"a}henb{\"u}hl}, {Kranich}, {La Barbera}, {Laille},
  {Leonardo}, {Lindfors}, {Lombardi}, {Longo}, {L{\'o}pez}, {Lorenz},
  {Majumdar}, {Maneva}, {Mankuzhiyil}, {Mannheim}, {Maraschi}, {Mariotti},
  {Mart{\'{\i}}nez}, {Mazin}, {Meucci}, {Miranda}, {Mirzoyan}, {Miyamoto},
  {Mold{\'o}n}, {Moles}, {Moralejo}, {Nieto}, {Nilsson}, {Ninkovic}, {Orito},
  {Oya}, {Paiano}, {Paoletti}, {Paredes}, {Partini}, {Pasanen}, {Pascoli},
  {Pauss}, {Pegna}, {Perez-Torres}, {Persic}, {Peruzzo}, {Prada}, {Prandini},
  {Puchades}, {Puljak}, {Reichardt}, {Rhode}, {Rib{\'o}}, {Rico}, {Rissi},
  {R{\"u}gamer}, {Saggion}, {Saito}, {Salvati}, {S{\'a}nchez-Conde},
  {Satalecka}, {Scalzotto}, {Scapin}, {Schultz}, {Schweizer}, {Shayduk},
  {Shore}, {Sierpowska-Bartosik}, {Sillanp{\"a}{\"a}}, {Sitarek}, {Sobczynska},
  {Spanier}, {Spiro}, {Stamerra}, {Steinke}, {Struebig}, {Suric}, {Takalo},
  {Tavecchio}, {Temnikov}, {Terzic}, {Tescaro}, {Teshima}, {Torres}, {Vankov},
  {Wagner}, {Zabalza}, {Zandanel}, {Zanin}, {Zapatero}, {MAGIC Collaboration},
  {Pfrommer}, {Pinzke}, {En{\ss}lin}, {Inoue}, \&
  {Ghisellini}}]{Aleksic:2010aa}
{Aleksi{\'c}}, J. {et~al.} 2010, \apj, 710, 634

\bibitem[{{Bell}(1978)}]{Bell:1978aa}
{Bell}, A.~R. 1978, \mnras, 182, 147

\bibitem[{{Berger} \& {Colella}(1989)}]{Berger:1989aa}
{Berger}, M.~J. \& {Colella}, P. 1989, J. Comp. Phys., 82, 64

\bibitem[{{Blandford} \& {Eichler}(1987)}]{Blandford:1987aa}
{Blandford}, R. \& {Eichler}, D. 1987, \physrep, 154, 1

\bibitem[{{Blandford} \& {Ostriker}(1978)}]{Blandford:1978aa}
{Blandford}, R.~D. \& {Ostriker}, J.~P. 1978, \apjl, 221, L29

\bibitem[{{Bonafede} {et~al.}(2009){Bonafede}, {Giovannini}, {Feretti},
  {Govoni}, \& {Murgia}}]{Bonafede:2009aa}
{Bonafede}, A., {Giovannini}, G., {Feretti}, L., {Govoni}, F., \& {Murgia}, M.
  2009, \aap, 494, 429

\bibitem[{{Bondi} \& {Hoyle}(1944)}]{Bondi:1944aa}
{Bondi}, H. \& {Hoyle}, F. 1944, \mnras, 104, 273

\bibitem[{{Brunetti} {et~al.}(2001){Brunetti}, {Setti}, {Feretti}, \&
  {Giovannini}}]{Brunetti:2001aa}
{Brunetti}, G., {Setti}, G., {Feretti}, L., \& {Giovannini}, G. 2001, \mnras,
  320, 365

\bibitem[{{Bryan} \& {Norman}(1997{\natexlab{a}})}]{Bryan:1997aa}
{Bryan}, G. \& {Norman}, M. 1997{\natexlab{a}}, 12th Kingston Meeting on
  Theoretical Astrophysics, proceedings of meeting held in Halifax; Nova
  Scotia; Canada October 17-19; 1996 (ASP Conference Series \# 123), ed.
  D.~Clarke. \& M.~Fall, 363

\bibitem[{{Bryan} \& {Norman}(1997{\natexlab{b}})}]{Bryan:1997ab}
---. 1997{\natexlab{b}}, Workshop on Structured Adaptive Mesh Refinement Grid
  Methods, ed. N.~Chrisochoides (IMA Volumes in Mathematics No. 117), 433

\bibitem[{{Bryan} {et~al.}(1995){Bryan}, {Norman}, {Stone}, {Cen}, \&
  {Ostriker}}]{Bryan:1995aa}
{Bryan}, G.~L., {Norman}, M.~L., {Stone}, J.~M., {Cen}, R., \& {Ostriker},
  J.~P. 1995, Computer Physics Communications, 89, 149

\bibitem[{{Burns} \& {LUNAR Consortium}(2009)}]{Burns:2009aa}
{Burns}, J.~O. \& {LUNAR Consortium}, t. 2009, ArXiv e-prints

\bibitem[{{Cassano} {et~al.}(2006){Cassano}, {Brunetti}, \&
  {Setti}}]{Cassano:2006aa}
{Cassano}, R., {Brunetti}, G., \& {Setti}, G. 2006, \mnras, 369, 1577

\bibitem[{{Clarke} \& {Ensslin}(2006)}]{Clarke:2006aa}
{Clarke}, T.~E. \& {Ensslin}, T.~A. 2006, \aj, 131, 2900

\bibitem[{{Clarke} {et~al.}(2001){Clarke}, {Kronberg}, \&
  {B{\"o}hringer}}]{Clarke:2001aa}
{Clarke}, T.~E., {Kronberg}, P.~P., \& {B{\"o}hringer}, H. 2001, \apjl, 547,
  L111

\bibitem[{{Drury}(1983)}]{Drury:1983aa}
{Drury}, L.~O. 1983, Reports on Progress in Physics, 46, 973

\bibitem[{{Efstathiou} {et~al.}(1985){Efstathiou}, {Davis}, {White}, \&
  {Frenk}}]{Efstathiou:1985aa}
{Efstathiou}, G., {Davis}, M., {White}, S.~D.~M., \& {Frenk}, C.~S. 1985,
  \apjs, 57, 241

\bibitem[{{Eisenstein} \& {Hu}(1999)}]{Eisenstein:1999aa}
{Eisenstein}, D.~J. \& {Hu}, W. 1999, \apj, 511, 5

\bibitem[{{Eisenstein} \& {Hut}(1998)}]{Eisenstein:1998aa}
{Eisenstein}, D.~J. \& {Hut}, P. 1998, \apj, 498, 137

\bibitem[{{Ensslin} {et~al.}(1998){Ensslin}, {Biermann}, {Klein}, \&
  {Kohle}}]{Ensslin:1998aa}
{Ensslin}, T.~A., {Biermann}, P.~L., {Klein}, U., \& {Kohle}, S. 1998, \aap,
  332, 395

\bibitem[{{Feretti}(2002)}]{Feretti:2002aa}
{Feretti}, L. 2002, in IAU Symposium, Vol. 199, The Universe at Low Radio
  Frequencies, ed. {A.~Pramesh Rao, G.~Swarup, \& Gopal-Krishna}, 133--+

\bibitem[{{Ferland} {et~al.}(1998){Ferland}, {Korista}, {Verner}, {Ferguson},
  {Kingdon}, \& {Verner}}]{Ferland:1998aa}
{Ferland}, G.~J., {Korista}, K.~T., {Verner}, D.~A., {Ferguson}, J.~W.,
  {Kingdon}, J.~B., \& {Verner}, E.~M. 1998, \pasp, 110, 761

\bibitem[{{Fusco-Femiano} {et~al.}(2001){Fusco-Femiano}, {Dal Fiume},
  {Orlandini}, {Brunetti}, {Feretti}, \& {Giovannini}}]{Fusco-Femiano:2001aa}
{Fusco-Femiano}, R., {Dal Fiume}, D., {Orlandini}, M., {Brunetti}, G.,
  {Feretti}, L., \& {Giovannini}, G. 2001, \apjl, 552, L97

\bibitem[{{Giacintucci} {et~al.}(2008){Giacintucci}, {Venturi}, {Macario},
  {Dallacasa}, {Brunetti}, {Markevitch}, {Cassano}, {Bardelli}, \&
  {Athreya}}]{Giacintucci:2008aa}
{Giacintucci}, S. {et~al.} 2008, ArXiv e-prints, 803

\bibitem[{{Govoni} {et~al.}(2006){Govoni}, {Murgia}, {Feretti}, {Giovannini},
  {Dolag}, \& {Taylor}}]{Govoni:2006aa}
{Govoni}, F., {Murgia}, M., {Feretti}, L., {Giovannini}, G., {Dolag}, K., \&
  {Taylor}, G.~B. 2006, \aap, 460, 425

\bibitem[{{Hockney} \& {Eastwood}(1988)}]{Hockney:1988aa}
{Hockney}, R.~W. \& {Eastwood}, J.~W. 1988, Computer Simulation Using Particles
  (Institute of Physics Publishing)

\bibitem[{{Hoeft} \& {Br{\"u}ggen}(2007)}]{Hoeft:2007aa}
{Hoeft}, M. \& {Br{\"u}ggen}, M. 2007, \mnras, 375, 77

\bibitem[{{Hoeft} {et~al.}(2008){Hoeft}, {Br{\"u}ggen}, {Yepes},
  {Gottl{\"o}ber}, \& {Schwope}}]{Hoeft:2008aa}
{Hoeft}, M., {Br{\"u}ggen}, M., {Yepes}, G., {Gottl{\"o}ber}, S., \& {Schwope},
  A. 2008, \mnras, 391, 1511

\bibitem[{{Jones} \& {Ellison}(1991)}]{Jones:1991aa}
{Jones}, F.~C. \& {Ellison}, D.~C. 1991, \ssr, 58, 259

\bibitem[{{Kang} {et~al.}(2007){Kang}, {Ryu}, {Cen}, \&
  {Ostriker}}]{Kang:2007ac}
{Kang}, H., {Ryu}, D., {Cen}, R., \& {Ostriker}, J.~P. 2007, \apj, 669, 729

\bibitem[{{Komatsu} {et~al.}(2008){Komatsu}, {Dunkley}, {Nolta}, {Bennett},
  {Gold}, {Hinshaw}, {Jarosik}, {Larson}, {Limon}, {Page}, {Spergel},
  {Halpern}, {Hill}, {Kogut}, {Meyer}, {Tucker}, {Weiland}, {Wollack}, \&
  {Wright}}]{Komatsu:2008aa}
{Komatsu}, E. {et~al.} 2008, ArXiv e-prints, 803

\bibitem[{{Miniati} {et~al.}(2001{\natexlab{a}}){Miniati}, {Jones}, {Kang}, \&
  {Ryu}}]{Miniati:2001ab}
{Miniati}, F., {Jones}, T.~W., {Kang}, H., \& {Ryu}, D. 2001{\natexlab{a}},
  \apj, 562, 233

\bibitem[{{Miniati} {et~al.}(2001{\natexlab{b}}){Miniati}, {Ryu}, {Kang}, \&
  {Jones}}]{Miniati:2001aa}
{Miniati}, F., {Ryu}, D., {Kang}, H., \& {Jones}, T.~W. 2001{\natexlab{b}},
  \apj, 559, 59

\bibitem[{{Miniati} {et~al.}(2000){Miniati}, {Ryu}, {Kang}, {Jones}, {Cen}, \&
  {Ostriker}}]{Miniati:2000aa}
{Miniati}, F., {Ryu}, D., {Kang}, H., {Jones}, T.~W., {Cen}, R., \& {Ostriker},
  J.~P. 2000, \apj, 542, 608

\bibitem[{{Napier}(2006)}]{Napier:2006aa}
{Napier}, P.~J. 2006, in Astronomical Society of the Pacific Conference Series,
  Vol. 356, Revealing the Molecular Universe: One Antenna is Never Enough, ed.
  {D.~C.~Backer, J.~M.~Moran, \& J.~L.~Turner}, 65--+

\bibitem[{{Norman} \& {Bryan}(1999)}]{Norman:1999aa}
{Norman}, M. \& {Bryan}, G. 1999, Numerical Astrophysics : Proceedings of the
  International Conference on Numerical Astrophysics 1998 (NAP98), held at the
  National Olympic Memorial Youth Center, Tokyo, Japan, March 10-13, 1998., ed.
  S.~M. {Miyama}, K.~{Tomisaka}, \& T.~{Hanawa} (Kluwer Academic), 19

\bibitem[{{Orr{\'u}} {et~al.}(2007){Orr{\'u}}, {Murgia}, {Feretti}, {Govoni},
  {Brunetti}, {Giovannini}, {Girardi}, \& {Setti}}]{Orru:2007aa}
{Orr{\'u}}, E. {et~al.} 2007, \aap, 467, 943

\bibitem[{{O'Shea} {et~al.}(2005{\natexlab{a}}){O'Shea}, {Bryan}, {Bordner},
  {Norman}, {Abel}, \& {Harkness}}]{OShea:2005aa}
{O'Shea}, B., {Bryan}, G., {Bordner}, J., {Norman}, M., {Abel}, T., \&
  {Harkness}, R. amd~{Kritsuk}, A. 2005{\natexlab{a}}, Adaptive Mesh Refinement
  - Theory and Applications, ed. T.~Plewa, T.~Linde, \& G.~Weirs
  (Springer-Verlag), 341

\bibitem[{{O'Shea} {et~al.}(2005{\natexlab{b}}){O'Shea}, {Nagamine},
  {Springel}, {Hernquist}, \& {Norman}}]{OShea:2005ab}
{O'Shea}, B.~W., {Nagamine}, K., {Springel}, V., {Hernquist}, L., \& {Norman},
  M.~L. 2005{\natexlab{b}}, \apjs, 160, 1

\bibitem[{{Paul} {et~al.}(2011){Paul}, {Iapichino}, {Miniati}, {Bagchi}, \&
  {Mannheim}}]{Paul:2011aa}
{Paul}, S., {Iapichino}, L., {Miniati}, F., {Bagchi}, J., \& {Mannheim}, K.
  2011, \apj, 726, 17

\bibitem[{{Pfrommer}(2008)}]{Pfrommer:2008ab}
{Pfrommer}, C. 2008, \mnras, 385, 1242

\bibitem[{{Pfrommer} {et~al.}(2008){Pfrommer}, {En{\ss}lin}, \&
  {Springel}}]{Pfrommer:2008aa}
{Pfrommer}, C., {En{\ss}lin}, T.~A., \& {Springel}, V. 2008, \mnras, 385, 1211

\bibitem[{{Pfrommer} {et~al.}(2006){Pfrommer}, {Springel}, {En{\ss}lin}, \&
  {Jubelgas}}]{Pfrommer:2006aa}
{Pfrommer}, C., {Springel}, V., {En{\ss}lin}, T.~A., \& {Jubelgas}, M. 2006,
  \mnras, 367, 113

\bibitem[{{Rephaeli} \& {Gruber}(2003)}]{Rephaeli:2003aa}
{Rephaeli}, Y. \& {Gruber}, D. 2003, \apj, 595, 137

\bibitem[{{Rephaeli} {et~al.}(2006){Rephaeli}, {Gruber}, \&
  {Arieli}}]{Rephaeli:2006aa}
{Rephaeli}, Y., {Gruber}, D., \& {Arieli}, Y. 2006, \apj, 649, 673

\bibitem[{{Roettiger} {et~al.}(1999){Roettiger}, {Stone}, \&
  {Burns}}]{Roettiger:1999aa}
{Roettiger}, K., {Stone}, J.~M., \& {Burns}, J.~O. 1999, \apj, 518, 594

\bibitem[{{Rottgering} {et~al.}(1997){Rottgering}, {Wieringa}, {Hunstead}, \&
  {Ekers}}]{Rottgering:1997aa}
{Rottgering}, H.~J.~A., {Wieringa}, M.~H., {Hunstead}, R.~W., \& {Ekers}, R.~D.
  1997, \mnras, 290, 577

\bibitem[{{Rudnick} {et~al.}(2009){Rudnick}, {Alexander}, {Andernach},
  {Battaglia}, {Brown}, {Brunetti}, {Burns}, {Clarke}, {Dolag}, {Farnsworth},
  {Giovannini}, {Hallman}, {Johnston-Hollitt}, {Jones}, {Kang}, {Kassim},
  {Kravtsov}, {Lazio}, {Lonsdale}, {McNamara}, {Myers}, {Owen}, {Pfrommer},
  {Ryu}, {Sarazin}, {Subrahmanyan}, {Taylor}, \& {Taylor}}]{Rudnick:2009aa}
{Rudnick}, L. {et~al.} 2009, in Astronomy, Vol. 2010, AGB Stars and Related
  Phenomenastro2010: The Astronomy and Astrophysics Decadal Survey, 253--+

\bibitem[{{Ryu} {et~al.}(2008){Ryu}, {Kang}, {Cho}, \& {Das}}]{Ryu:2008aa}
{Ryu}, D., {Kang}, H., {Cho}, J., \& {Das}, S. 2008, Science, 320, 909

\bibitem[{{Ryu} {et~al.}(2003){Ryu}, {Kang}, {Hallman}, \&
  {Jones}}]{Ryu:2003aa}
{Ryu}, D., {Kang}, H., {Hallman}, E., \& {Jones}, T. 2003, \apj, 593, 599

\bibitem[{{Skillman} {et~al.}(2008){Skillman}, {O'Shea}, {Hallman}, {Burns}, \&
  {Norman}}]{Skillman:2008aa}
{Skillman}, S., {O'Shea}, B., {Hallman}, E., {Burns}, J., \& {Norman}, M. 2008,
  \apj, 689, 1063

\bibitem[{{Smith} {et~al.}(2008){Smith}, {Sigurdsson}, \&
  {Abel}}]{Smith:2008aa}
{Smith}, B., {Sigurdsson}, S., \& {Abel}, T. 2008, \mnras, 385, 1443

\bibitem[{{Spitkovsky}(2008)}]{Spitkovsky:2008aa}
{Spitkovsky}, A. 2008, \apjl, 682, L5

\bibitem[{{Stone} \& {Norman}(1992{\natexlab{a}})}]{Stone:1992aa}
{Stone}, J.~M. \& {Norman}, M.~L. 1992{\natexlab{a}}, \apjs, 80, 753

\bibitem[{{Stone} \& {Norman}(1992{\natexlab{b}})}]{Stone:1992ab}
---. 1992{\natexlab{b}}, \apjs, 80, 791

\bibitem[{{Turk} {et~al.}(2011){Turk}, {Smith}, {Oishi}, {Skory}, {Skillman},
  {Abel}, \& {Norman}}]{Turk:2011aa}
{Turk}, M.~J., {Smith}, B.~D., {Oishi}, J.~S., {Skory}, S., {Skillman}, S.~W.,
  {Abel}, T., \& {Norman}, M.~L. 2011, \apjs, 192, 9

\bibitem[{{van Weeren} {et~al.}(2009{\natexlab{a}}){van Weeren}, {Rottgering},
  {Bagchi}, {Raychaudhury}, {Intema}, {Miniati}, {Ensslin}, {Markevitch}, \&
  {Erben}}]{van-Weeren:2009ab}
{van Weeren}, R.~J. {et~al.} 2009{\natexlab{a}}, ArXiv e-prints

\bibitem[{{van Weeren} {et~al.}(2009{\natexlab{b}}){van Weeren}, {Rottgering},
  {Bruggen}, \& {Cohen}}]{van-Weeren:2009aa}
{van Weeren}, R.~J., {Rottgering}, H.~J.~A., {Bruggen}, M., \& {Cohen}, A.
  2009{\natexlab{b}}, ArXiv e-prints

\bibitem[{{Vazza} {et~al.}(2009){Vazza}, {Brunetti}, \&
  {Gheller}}]{Vazza:2009aa}
{Vazza}, F., {Brunetti}, G., \& {Gheller}, C. 2009, \mnras, 395, 1333

\bibitem[{{Warren} {et~al.}(2006){Warren}, {Abazajian}, {Holz}, \&
  {Teodoro}}]{Warren:2006aa}
{Warren}, M.~S., {Abazajian}, K., {Holz}, D.~E., \& {Teodoro}, L. 2006, \apj,
  646, 881

\bibitem[{{Woodward} \& {Colella}(1984)}]{Woodward:1984aa}
{Woodward}, P.~R. \& {Colella}, P. 1984, J. Comp. Phys., 54, 174

\end{thebibliography}
\end{document}